\documentclass[10pt,journal,epsfig]{IEEEtran}
\usepackage{graphicx}
\usepackage{subfigure,color}
\usepackage{amssymb}
\usepackage{cite,comment}
\usepackage{amsmath,setspace}
\usepackage{bm,comment,epstopdf}
\usepackage{algorithm}
\usepackage{algpseudocode}
\makeatletter

\newcommand{\Rmnum}[1]{\expandafter\@slowromancap\romannumeral #1@}
\newcommand{\mv}[1]{\mbox{\boldmath{$ #1 $}}}

\newcommand{\tabincell}[2]{\begin{tabular}{@{}#1@{}}#2\end{tabular}}
\makeatother

\newtheorem{proposition}{Proposition}

\newtheorem{definition}{Definition}

\begin{document}
\title{Joint Beam Routing and Resource Allocation Optimization for Multi-IRS-Reflection Wireless Power Transfer}
\author{Weidong Mei, \IEEEmembership{Member, IEEE}, Zhi Chen, \IEEEmembership{Senior Member, IEEE}, and Rui Zhang, \IEEEmembership{Fellow, IEEE}
\thanks{W. Mei and Z. Chen are with the National Key Laboratory of Wireless Communications, University of Electronic Science and Technology of China, Chengdu 611731, China (e-mail: wmei@uestc.edu.cn, chenzhi@uestc.edu.cn).}
\thanks{R. Zhang is with School of Science and Engineering, Shenzhen Research Institute of Big Data, The Chinese University of Hong Kong, Shenzhen, Guangdong 518172, China (e-mail: rzhang@cuhk.edu.cn). He is also with the Department of Electrical and Computer Engineering, National University of Singapore, Singapore 117583 (e-mail: elezhang@nus.edu.sg).}}
\maketitle

\begin{abstract}
Intelligent reflecting surface (IRS) can be densely deployed in complex environments to create cascaded line-of-sight (LoS) links between base stations (BSs) and users, which significantly enhance the signal coverage. In this paper, we consider the wireless power transfer (WPT) from a multi-antenna BS to multiple energy users (EUs) by exploiting the signal beam routing via multi-IRS reflections. First, we present a baseline beam routing scheme with each IRS serving at most one EU, where the BS transmits wireless power to all EUs simultaneously while the signals to different EUs undergo disjoint sets of multi-IRS reflection paths. Under this setup, we aim to tackle the joint beam routing and resource allocation optimization problem by jointly optimizing the reflection paths for all EUs, the active/passive beamforming at the BS/each involved IRS, as well as the BS's power allocation for different EUs to maximize the minimum received signal power among all EUs. Next, to further improve the WPT performance, we propose two new beam routing schemes, namely dynamic beam routing and subsurface-based beam routing, where each IRS can serve multiple EUs via different time slots and different subsurfaces, respectively. In particular, we prove that dynamic beam routing outperforms subsurface-based beam routing in terms of minimum harvested power among all EUs. In addition, we show that the optimal performance of dynamic beam routing is achieved by assigning all EUs with orthogonal time slots for WPT. A clique-based optimization approach is also proposed to solve the joint beam routing and resource allocation problems for the baseline beam routing and proposed dynamic beam routing schemes. Numerical results are finally presented, which demonstrate the superior performance of the proposed dynamic beam routing scheme to the baseline scheme.
\end{abstract}

\begingroup
\allowdisplaybreaks
\section{Introduction}
In recent years, intelligent reflecting surface (IRS) and its equivalents (e.g., reconfigurable intelligent surface (RIS)) have emerged as a cost-effective solution to enhance the spectrum and energy efficiency of wireless networks. By independently tuning the phase shifts of massive low-cost reflecting elements, IRS can dynamically reshape wireless propagation channels to achieve favorable signal transmissions\cite{wu2019towards,basar2019wireless}. Moreover, different from traditional active base stations (BSs) or relays, IRS reflecting elements dispense with transmit/receive radio frequency (RF) chains, thus dramatically reducing the hardware cost and energy consumption\cite{wu2019towards,basar2019wireless}. Due to its many appealing advantages, there have been extensive research works on the design and performance analysis of IRSs in different system setups and various application scenarios, by addressing its main practical challenges including reflection design, channel acquisition, deployment optimization and so on (see e.g., \cite{wu2019towards,basar2019wireless,wu2020intelligent,di2020smart,yuan2021reconfigurable,bjornson2022reconfigurable,pan2022overview,zheng2022survey,swindlehurst2022channel,you2022deploy} and the references therein).

\begin{figure}[!t]
\centering
\includegraphics[width=3.4in]{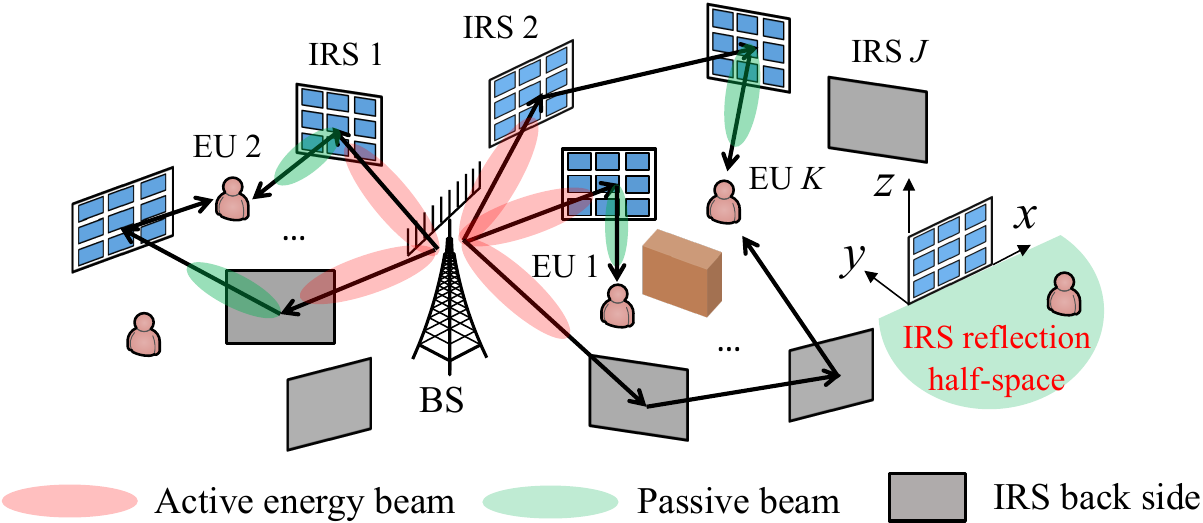}
\DeclareGraphicsExtensions.
\caption{Multi-IRS-reflection aided wireless power transfer.}\label{SysModel}
\vspace{-9pt}
\end{figure}
However, most existing works on IRS have considered one single IRS or multiple IRSs each reflecting the signal from/to the BS once, but ignored the more general multi-IRS-reflection links, which can also be utilized to further improve the end-to-end channel condition. For example, in complex environment with dense obstacles, multi-IRS reflections can create more available cascaded line-of-sight (LoS) links to bypass the obstacles between the BS and remote user locations, thereby significantly enhancing the BS's signal coverage\cite{mei2022intelligent}. In addition, more pronounced cooperative passive beamforming (CPB) gain can be reaped from the multi-IRS-reflection LoS links, which helps compensate for the more severe multiplicative path loss among cascaded IRSs\cite{mei2022intelligent}. Due to the above reasons, the strength of a properly designed multi-IRS-reflection LoS link can even be higher than that of any constituent single-IRS-reflection links\cite{mei2022intelligent}. Inspired by this, recent works have investigated the multi-IRS reflection design, channel acquisition, and deployment issues in large-scale wireless networks, which, however, are generally more challenging than their counterpart issues in single-IRS-reflection systems, due to the more complicated multi-IRS signal reflections involved.

First, in the existing works on {\it multi-IRS reflection design}, the authors in \cite{huang2021multi} and \cite{liang2022multi} proposed a deep reinforcement learning (DRL) approach and a channel decomposition approach to jointly optimize the BS/IRS active/passive beamforming in a given multi-IRS-reflection link, respectively. To maximally leverage the multi-IRS LoS path diversity gain, the authors in \cite{mei2020cooperative,mei2022mbmh} studied a new passive beam routing design problem, with the goal to select the optimal multi-IRS-reflection path for each user and jointly optimize the BS/IRS active/passive beamforming in each selected path, such that the end-to-end BS-user channel power gain is maximized. A more general multi-path beam routing scheme was also proposed in \cite{mei2022split} to improve over its single-path counterpart proposed in \cite{mei2020cooperative} and \cite{mei2022mbmh}, where multiple reflection paths are selected for a user, with the signals over them coherently combined at the user's receiver. The authors in \cite{zhang2022multi} delved into the beam routing design arising in a new hybrid multi-IRS-reflection system with one active IRS and multiple passive IRSs. Furthermore, the authors in \cite{bhowal2022mimo,tyrovolas2022performance,zhakipov2023accurate,wang2022novel,liu2022intelligent} analyzed different performance metrics for the multi-IRS-reflection system, e.g., ergodic capacity, outage probability, and error rate, and derived their tractable bounds. Second, for {\it multi-IRS channel acquisition}, the authors in \cite{mei2021distributed} proposed a distributed beam training scheme with combined online and offline processing, which avoids the prohibitive time overhead due to multi-IRS-reflection channel estimation and beam training in real time. Moreover, the authors in \cite{xu2023coordinating} proposed a blind beamforming method, where the multi-IRS CPB in any multi-reflection link is designed based on only the power measurements at each user for a sequence of randomly generated phase shifts across its serving IRSs. It was shown in \cite{xu2023coordinating} that their proposed blind method can achieve the same CPB gain as that achievable under perfect channel state information (CSI) in \cite{mei2020cooperative}. Last but not least, for {\it multi-IRS deployment}, a graph-based system model and optimization framework was proposed in \cite{mei2023deploy} for joint BS and IRS deployment optimization in a given region, where a fundamental performance-cost trade-off was unveiled. 

It is worth noting that the above works on multi-IRS reflection all considered communication systems and assumed that each IRS only serves at most one user, while exploiting the potential of each IRS to serve multiple users may further improve the system performance. This thus motivates the current paper to investigate the new setup of multi-IRS-reflection aided wireless power transfer (WPT) where each IRS can potentially serve multiple users by different means. Specifically, we consider a WPT system where a multi-antenna BS intends to send wireless power to multiple energy users (EUs) in the presence of multiple helping IRSs at known locations, as shown in Fig.\,\ref{SysModel}. The main contributions of this paper are summarized as follows.
\begin{itemize}
	\item First, we present a baseline beam routing scheme with each IRS serving at most one EU, where the BS wirelessly transfers power to all EUs simultaneously via disjoint sets of multi-IRS reflection paths. Note that this scheme extends our proposed single-user multi-path beam routing scheme in \cite{mei2022split} to the general multi-user case. Specifically, we aim to jointly optimize the set of IRS reflection paths selected for each EU, the active/passive beamforming at the BS/each selected IRS, as well as the BS's power allocation for different EUs to maximize the minimum received signal power among all EUs. To avoid the undesired randomly scattered interference among different selected reflection paths for different EUs, we propose a new method by adding randomly generated common phase shifts to all the reflecting elements of some IRSs without degrading their beamforming gain, and show that the effect of random IRS scattering can be averaged out over time at each EU's receiver and thus becomes negligible.
	\item Next, to further improve the WPT performance over the baseline scheme, we propose two new beam routing schemes, namely, dynamic beam routing and subsurface-based beam routing, whereby each IRS can serve multiple EUs over different time slots and via different subsurfaces, respectively. In particular, we prove that dynamic beam routing always yields a better performance than subsurface-based beam routing in terms of minimum harvested power among EUs. In addition, we show that the optimal performance of dynamic beam routing is achieved by assigning all EUs with orthogonal time slots for WPT. A clique-based optimization approach is also proposed to solve the joint beam routing and resource allocation problems for the baseline beam routing and proposed dynamic beam routing schemes. Numerical results demonstrate that the proposed dynamic beam routing can significantly outperform the baseline (static) beam routing and unveil useful insights.
\end{itemize}

The rest of this paper is organized as follows. Section \Rmnum{2} presents the system model of the multi-IRS-reflection aided WPT. Section \Rmnum{3} presents the baseline beam routing scheme and its associated joint beam routing and resource allocation optimization problem. Section \Rmnum{4} presents the proposed dynamic and subsurface-based beam routing schemes and compares their performance analytically. Section \Rmnum{5} presents the joint beam routing and resource allocation optimization problem for the dynamic beam routing and the proposed solutions to this problem and that for the baseline scheme. Section \Rmnum{6} presents the simulation results to show the efficacy of the proposed scheme. Finally, Section \Rmnum{7} concludes this paper and discusses future work.

The following notations are used in this paper. Bold symbols in capital letter and small letter denote matrices and vectors, respectively. The conjugate, transpose and conjugate transpose of a vector or matrix are denoted as ${(\cdot)}^{*}$, ${(\cdot)}^{T}$ and ${(\cdot)}^{H}$, respectively. ${\mathbb{R}}^n$ (${\mathbb{C}}^n$) denotes the set of real (complex) vectors of length $n$. For a complex number $s$, $s^*$ and $\lvert s \rvert$ denote its conjugate and amplitude, respectively. For a vector ${\mv a} \in {\mathbb{C}}^n$, ${\rm diag}({\mv a})$ denotes an $n \times n$ diagonal matrix whose entries are given by the elements of $\mv a$; while for a square matrix ${\mv A} \in {\mathbb{C}}^{n \times n}$, ${\rm diag}({\mv A})$ denotes an $n \times 1$ vector that contains the $n$ diagonal elements of ${\mv A}$. $j$ denotes the imaginary unit, i.e., $j^2=-1$. For two sets $A$ and $B$, $A \cup B$ denotes the union of $A$ and $B$. $\emptyset$ denotes an empty set. $\otimes$ denotes the Kronecker product. For ease of reference, the main symbols used in this paper are listed in Table \ref{variable}.

\begin{table*}[htbp]
\centering
\caption{List of Main Symbols}\label{variable}
\resizebox{0.8\textwidth}{!}{
\begin{tabular}{|c|l|c|l|}
\hline
{\bf{Symbol}} & {\bf{Description}} & {\bf{Symbol}} & {\bf{Description}}\\
\hline
$N_B$ & Number of BS antennas & $K$ & Number of EUs \\
\hline
$\cal K$ & Set of EUs & $J$ & Number of IRSs \\
\hline
${\cal J}$ & Set of IRSs & $M$ & Number of reflecting elements per IRS \\
\hline
${\mv w}_B$ & Energy beamforming vector of the BS & ${\cal W}_B$ & Beamforming codebook of the BS \\
\hline
${\mv \Phi}_j$ & Reflection matrix of IRS $j$ & ${\mv \theta}_j$ & Passive beamforming vector of IRS $j$ \\
\hline
${\cal W}_I$ & Beamforming codebook of the IRSs & $d_{i,j}$ & Distance between nodes $i$ and $j$\\
\hline
$G_L$ & LoS graph of the system & $V_L/E_L$ & Vertex/edge set of $G_L$\\
\hline
$e_{i,j}$ & edge from vertex $i$ to vertex $j$ in $G_L$ & ${\mv H}_{0,j}$ & Channel from the BS to IRS $j$\\
\hline
${\mv S}_{i,j}$ & Channel from IRS $i$ to IRS $j$ & ${\mv g}_{j,J+k}^{H}$ & Channel from IRS $j$ to EU $k$\\
\hline
$h_{0,J+k}(\Omega)$ & BS-EU $k$ channel over a path $\Omega$ & $\kappa_{0,J+k}(\Omega)$ & BS-EU $k$ LoS gain over a path $\Omega$\\
\hline
$D_{0,J+k}(\Omega)$ & BS-EU $k$ distance over a path $\Omega$ & $\tilde \Gamma_k$ & Set of all LoS paths from the BS to EU $k$\\
\hline
$H_{0,J+k}(\Omega)$ & \tabincell{l}{Maximum BS-EU $k$ channel power gain\\ over a path $\Omega$} & ${\mv w}_B(j)$ & \tabincell{l}{Optimal BS beamforming vector to\\ transmit to IRS $j$}\\
\hline
${\mv\theta}(i,j,r)$ & \tabincell{l}{Optimal passive beamforming vector for\\ IRS $j$ to reflect from node $i$ to node $r$} & $\Gamma_k$ & Set of selected paths for EU $k$, $\Gamma_k \subseteq \tilde \Gamma_k$\\
\hline
$\Omega_k^{(q)}$ & $q$-th reflection path in $\Gamma_k$ & $L^{(q)}_k$ & Number of IRSs in $\Omega_k^{(q)}$\\
\hline
$a^{(q)}_{k,l}$ & Index of the $l$-th IRS in $\Omega_k^{(q)}$ & ${\cal N}_0$ & Set of all first reflecting IRSs\\
\hline
$\alpha_j$ & Total transmit power allocated to ${\mv w}_B(j)$ & $\rho_k$ & Total power allocated to EU $k$\\
\hline
$\tilde E_{k,\max}$ & Combined received signal power at EU $k$ & $\omega_j$ & Common phase shift appended to IRS $j$\\
\hline
$\omega(\Omega)$ & \tabincell{l}{Overall common phase shift appended to\\ a path $\Omega$} & $T$ & \tabincell{l}{Total number of time slots in dynamic\\ beam routing}\\
\hline
$\tau_t$ & Length of time slot $t$ & $\Gamma_{k,t}$ & Set of paths for EU $k$ in time slot $t$\\
\hline
$\rho_{k,t}$ & \tabincell{l}{Total power allocated to EU $k$ in time\\ slot $t$} & $E_{k,\text{D}}$ & \tabincell{l}{Average received power by EU $k$ with\\ dynamic beam routing}\\
\hline
$\mu_{j,k}$ & \tabincell{l}{Fraction of the elements of IRS $j$\\ allocated to EU $k$} & $j_k$ & \tabincell{l}{Index of the subsurface of IRS $j$\\ associated with EU $k$}\\
\hline
${\cal J}_k$ & Set of subsurfaces serving EU $k$ & $\Upsilon_k$ & Set of subsurface paths selected for EU $k$\\
\hline
$B^{(r)}_k$ & $r$-th subsurface path in $\Upsilon_k$ & $E_{k,\text{S}}$ & \tabincell{l}{Received signal power by EU $k$ with\\ subsurface-based beam routing}\\
\hline
$\eta_{k,t}$ & Auxiliary variable, $\eta_{k,t}=\rho_{k,t}\tau_t$ & $U$ & Number of candidate paths for each EU\\
\hline
${\cal U}_k$ & Set of candidate paths for EU $k$ & $\cal U$ & Set of candidate paths for all EUs\\
\hline
$P_k^{(u)}$ & $u$-th candidate path in ${\cal U}_k$ & $G_{p,k}$ & Path graph generated based on ${\cal U}_k$\\
\hline
$G_p$ & Path graph generated based on ${\cal U}$ & $v(P_k^{(u)})$ & Corresponding vertex of $P_k^{(u)}$ in $G_p$ or $G_{p,k}$\\
\hline
$C^{(k)}$ & \tabincell{l}{Reflection paths for EU $k$ specified by a\\ clique $C$} & \tabincell{l}{${\cal C}_{k,\max}$\\(${\cal C}_{\max}$)} & Set of maximal cliques in $G_{p,k}$ ($G_p$)\\
\hline
\end{tabular}}
\end{table*}

\section{System Model}
As shown in Fig.\,\ref{SysModel}, we consider a multi-IRS-aided wireless system where multiple distributed IRSs are employed to assist in the WPT from a BS with $N_B$ antennas to $K$ single-antenna EUs\footnote{Although our considered IRS-aided WPT shares certain similarity with the IRS-aided wireless information transfer (WIT), they also differ in terms of design objective and practical considerations/constraints\cite{wu2022intelligent}.}, denoted by ${\cal K}\triangleq \{1,2,\cdots,K\}$. We assume that there are $J$ IRSs in total, denoted by ${\cal J}\triangleq \{1,2,\cdots,J\}$, and each IRS is equipped with $M$ reflecting elements. For convenience, we label the BS, IRS $j, j \in {\cal J}$, and EU $k, k \in {\cal K}$ as nodes $0$, $j$, and $J+k$, respectively. Due to the dense obstacles in the environment, the BS can only transfer the power to each EU via one or multiple IRS reflection paths between them, each formed by a set of IRSs which successively reflect the signal from the BS to each EU.

Let ${\mv w}_B \in {\mathbb C}^{N_B \times 1}$ denote the (active) energy beamforming vector at the BS. To ease practical implementation, we consider that the BS employs a prior-designed beamforming codebook consisting of orthogonal and unit-power beams, denoted as ${\cal W}_B$, i.e., ${\mv w}_B \in {\cal W}_B$. Let ${\mv \Phi}_j={\rm diag}({\mv \theta}_j) \in {\mathbb C}^{M \times M}$ denote the reflection matrix of IRS $j, j \in \cal J$, where ${\mv \theta}_j \in {\mathbb C}^{M \times 1}$ denotes its passive beamforming vector and is assumed to be selected from a passive beamforming codebook ${\cal W}_I$, i.e., ${\mv \theta}_j \in {\cal W}_I, j \in \cal J$. It is also assumed that the BS and each IRS are equipped with a uniform linear array (ULA) and a uniform rectangular array (URA), respectively. The antenna and element spacing at the BS and each IRS is denoted as $d_A$ and $d_I$, respectively. The distance between any two nodes $i$ and $j$ in the system is denoted as $d_{i,j}$, with a reference element selected at the BS/each IRS. Since each IRS can only achieve 180$^\circ$ half-space reflection, for effective signal reflection from the BS to any EU via a reflection path, the previous and next nodes of each IRS in this path should be located in its reflection half-space. Moreover, we only consider the reflection paths where the energy signal is reflected outward from an IRS (e.g., IRS $j$) to a farther IRS from the BS (e.g., IRS $i$ with $d_{0,i}>d_{0,j}$) in our design, but ignore the links in the opposite direction as they typically incur more severe path loss\cite{mei2022intelligent}.

To enhance the strength of all BS-EU reflection links and leverage the rich LoS path diversity in the multi-IRS system, we consider that only the LoS links in the system are leveraged for dedicated WPT, while treating all non-LoS (NLoS) links as part of environment scattering, which generally have only marginal effect on the received signal power at each EU\cite{mei2022mbmh,mei2021distributed}. Accordingly, to depict all LoS paths from the BS to all EUs efficiently, we define an LoS graph for all nodes in the system, denoted as $G_L=(V_L,E_L)$, where $V_L=\{0,1,2,\cdots,J+1\}$ and $E_L$ denote the sets of vertices and edges in $G_L$, respectively. In particular, there exists an edge from vertex $i$ to vertex $j$, denoted as $e_{i,j}$, if there is an LoS path between nodes $i$ and $j$ and the following two conditions are satisfied: 1) effective signal reflection can be achieved from node $i$ to node $j$ subject to the half-space reflection constraint at any IRS involved (i.e., node $i$ and/or $j$); and 2) node $j$ is farther away from the BS than node $i$, i.e., $d_{0,j} > d_{0,i}$, to reflect the signal outward (except that node $j$ is an EU). Based on the above, each LoS path from the BS to each EU $k, k \in \cal K$ corresponds to a path from vertex 0 to vertex $J+k$ in $G_L$. In this paper, we assume that the LoS graph $G_L$ is known at the BS via separate channel modeling/measurement\cite{mei2022intelligent}.

Let ${\mv H}_{0,j} \in {\mathbb C}^{M \times N}, j \in {\cal J}$ denote the channel from the BS to IRS $j$, ${\mv S}_{i,j} \in {\mathbb C}^{M \times M}, i,j \in {\cal J}, i \ne j$ denote that from IRS $i$ to IRS $j$, and ${\mv g}_{j,J+k}^{H} \in {\mathbb C}^{1 \times M}, j \in {\cal J}$ denote that from IRS $j$ to EU $k$. We assume that the BS and all IRSs are properly deployed, such that if there exists an LoS path from node $i$ (BS/IRS) to node $j$ (IRS/EU), the signal from node $i$ and that to node $j$ can be well approximated as a uniform plane wave. As such, the LoS channel from node $i$ to node $j$ (if any) can be modeled as the product of steering vectors at the two sides. Specifically, let ${\mv e}(\phi,N)=[1,e^{-j\pi\phi},\cdots,e^{-j\pi(N-1)\phi}]^T \in {\mathbb C}^{N \times 1}$ denote the steering vector of a ULA equipped with $N$ elements, where $\phi$ denotes the phase difference between its two adjacent elements. Then, the transmit array response at the BS with respect to (w.r.t.) node $j$ (IRS) is expressed as
\begin{equation}\label{array1}
	{\mv a}_B(\vartheta_{0,j})={\mv e}\Big(\frac{2d_A}{\lambda}\cos \vartheta_{0,j},N_B\Big),
\end{equation}
where $\lambda$ is the carrier wavelength, and $\vartheta_{0,j}$ denotes the angle-of-departure (AoD) from the BS to node $j$.

While for the URA at each IRS, we establish a local coordinate system for it and assume that the URA is parallel to the $x$-$z$ plane (see Fig.\,\ref{SysModel}) . As such, the transmit array response of each IRS $j$ w.r.t. node $i, i \ne j$ (IRS or EU) is expressed as the Kronecker product of two steering vector functions in the horizontal and vertical directions, respectively, i.e.,
\begin{align}
{\mv a}_I(\vartheta^a_{j,i},\vartheta^e_{j,i})={\mv e}\Big(\frac{2d_I}{\lambda}\sin\vartheta^e_{j,i}\cos&\vartheta^a_{j,i},M_1\Big) \otimes\nonumber\\
& {\mv e}\Big(\frac{2d_I}{\lambda}\cos\vartheta^e_{j,i},M_2\Big),\label{array2}
\end{align}
where $\vartheta^e_{j,i}$ and $\vartheta^a_{j,i}$ denote the elevation and azimuth AoDs from IRS $j$ to node $i$, respectively. Similarly, define $\varphi^a_{j,i}$/$\varphi^e_{j,i}$ as the azimuth/elevation angle-of-arrival (AoA) at IRS $j$ from node $i$ (BS or IRS). Then, its receive array response w.r.t. node $i$ can be obtained by replacing $\vartheta^a_{j,i}$ and $\vartheta^e_{j,i}$ in (\ref{array2}) with $\varphi^a_{j,i}$ and $\varphi^e_{j,i}$, respectively, i.e., ${\mv a}_I(\varphi^a_{j,i},\varphi^e_{j,i})$. 

Based on the above, if $e_{0,j} \in E_L$, the BS-IRS $j$ LoS channel is expressed as
\begin{equation}\label{Ch1}
{\mv H}_{0,j} = \frac{\sqrt \beta}{d_{0,j}}e^{-\frac{j2\pi d_{0,j}}{\lambda}}\underbrace{{\mv a}_I(\varphi^a_{j,0},\varphi^e_{j,0})}_{\triangleq{\tilde{\mv h}}_{j,2}}\underbrace{{\mv a}^H_B(\vartheta_{0,j})}_{\triangleq{\tilde{\mv h}}_{j,1}}, \;j \in {\cal J},
\end{equation}
where $\beta\, (<1)$ denotes the LoS path gain at the reference distance of 1 meter (m). Similarly, if $e_{i,j} \in E_L, i,j \in \cal J$, the IRS $i$-IRS $j$ channel is given by
\begingroup\makeatletter\def\f@size{9.4}\check@mathfonts
\def\maketag@@@#1{\hbox{\m@th\normalsize\normalfont#1}}%
\begin{equation}\label{Ch2}
{\mv S}_{i,j} \!\!=\!\! \frac{\sqrt \beta}{d_{i,j}}e^{-\frac{j2\pi d_{i,j}}{\lambda}}\underbrace{{\mv a}_I(\varphi^a_{j,i},\varphi^e_{j,i})}_{\triangleq {\tilde{\mv s}}_{i,j,2}}\underbrace{{\mv a}_I(\vartheta^a_{i,j},\vartheta^e_{i,j})}_{\triangleq {\tilde{\mv s}}_{i,j,1}}, \;i, j \in {\cal J}, i \ne j.
\end{equation}\endgroup
Finally, if $e_{j,J+k} \in E_L$, the IRS $j$-EU $k$ channel is expressed as
\begin{equation}\label{Ch3}
{\mv g}^H_{j,J+k} \!=\! \frac{\sqrt \beta}{d_{j,J+k}}e^{-\frac{j2\pi d_{{j,J+k}}}{\lambda}}\underbrace{{\mv a}_I(\vartheta^a_{j,J+k},\vartheta^e_{j,J+k})}_{\triangleq {\tilde{\mv g}}_{j,J+k}}, \;j \!\in\! {\cal J}, k \!\in\! {\cal K}.
\end{equation}

Based on (\ref{Ch1})-(\ref{Ch3}), we can characterize the end-to-end channel from the BS to EU $k, k \in \cal K$ in any given multi-reflection LoS path, denoted as $\Omega=\{a_1,a_2,\cdots,a_L\}$ with $e_{a_l,a_{l+1}} \in E_L, l=0,1,\cdots,L$, where $L \ge 1$ and $a_l \in \cal J$ denote the number of IRSs in $\Omega$ and the index of the $l$-th IRS, respectively, and we set $a_0=0$ and $a_{L+1}=J+k$. Then, the BS-EU $k$ end-to-end channel for a given $\Omega$ is expressed as
\begin{align}
h_{0,J+k}(\Omega)&={\mv g}^H_{a_L,J+k}{\mv \Phi}_{a_L}\Big(\prod\limits_{l=1}^{L-1}{\mv S}_{a_l,a_{l+1}}{\mv \Phi}_{a_l}\Big){\mv H}_{0,a_1}{\mv w}_B \nonumber\\
&=\kappa_{0,J+k}(\Omega)e^{-j\frac{2\pi D_{0,J+k}(\Omega)}{\lambda}}\Big(\prod\limits_{l=1}^{L}A_l\Big)({\tilde{\mv h}}^H_{a_1,1}{\mv w}_B),\label{e2eCh1}
\end{align}
where $\kappa_{0,J+k}(\Omega)=(\sqrt\beta)^{L+1}\prod\nolimits_{l=0}^{L}d^{-1}_{a_l,a_{l+1}}$ and $D_{0,J+k}(\Omega)=\sum\nolimits_{l=0}^{L}d_{a_l,a_{l+1}}$ denote the end-to-end path gain and distance over $\Omega$, respectively, and
\begin{equation}\label{Al}
A_l = \begin{cases}
	{\tilde{\mv s}}^H_{a_1,a_2,1}{\mv \Phi}_{a_1}{\tilde{\mv h}}_{a_1,2} &{\text{if}}\;\;l=1\\
	\tilde{\mv g}^H_{a_L,J+k}{\mv \Phi}_{a_L}{\tilde{\mv s}}_{a_{L-1},a_L,2} &{\text{if}}\;\;l=L\\
	{\tilde{\mv s}}^H_{a_l,a_{l+1},1}{\mv \Phi}_{a_l}{\tilde{\mv s}}_{a_{l-1},a_l,2} &{\text{otherwise}}.
	\end{cases}
\end{equation}

Let $\tilde\Gamma_k$ denote the set of all LoS paths from the BS to EU $k$ in our considered system. Then, their overall end-to-end channel is expressed as
\begin{equation}\label{e2eCh2}
h_{0,J+k}=\sum\limits_{\Omega \in \tilde\Gamma_k}h_{0,J+k}(\Omega), k \in {\cal K}.
\end{equation}

However, based on (\ref{e2eCh2}), it is generally difficult to derive the optimal BS and IRS beamforming vectors for WPT, since they are intricately coupled with each other in different paths in $\tilde\Gamma_k, k \in \cal K$. To tackle this challenge, by leveraging the rich LoS path diversity in the multi-IRS-aided wireless network, we first consider a baseline multi-user multi-path beam routing scheme in Section \ref{conv},  where a set of separated LoS reflection paths are selected for each EU as well as for different EUs, as shown in Fig.\,\ref{SysModel}, thereby decoupling the BS and IRS beamforming design over different paths. However, in this scheme, each IRS only serves at most one EU, which may result in suboptimal WPT performance. Hence, in Section \ref{proposed}, we further propose two new beam routing schemes to improve the WPT performance, where each IRS can serve multiple EUs over different time slots and via different subsurfaces, respectively.

\section{Beam Routing with Each IRS Serving a Single EU}\label{conv}
In this section, we present the baseline beam routing design for WPT, with each IRS serving at most one EU.

\subsection{Optimal BS and IRS Beamforming Design}
First, it is noted that to maximize the end-to-end channel power gain over a single path (e.g., $\lvert h_{0,J+k}(\Omega) \rvert^2$ in (\ref{e2eCh1})), the optimal active beamforming ${\mv w}_B$ and passive beamforming at each IRS $a_l$, ${\mv \theta}_{a_l}$, should be set as\cite{mei2020cooperative}
\begingroup\makeatletter\def\f@size{9.2}\check@mathfonts
\def\maketag@@@#1{\hbox{\m@th\normalsize\normalfont#1}}%
\begin{align}
	{\mv w}^{\star}_B&={\mv w}_B(a_1) \triangleq \arg \mathop{\max}\limits_{{\mv w} \in {\cal W}_B} \lvert{\tilde{\mv h}}^H_{a_1,1}\mv w \rvert, \label{abf}\\
	{\mv \theta}_{a_l}^{\star}&={\mv \theta}(a_{l-1},a_l,a_{l+1})\nonumber\\
	&\triangleq
\begin{cases}
	\arg \mathop{\max}\limits_{{\mv \theta} \in {\cal W}_I}\lvert{\tilde{\mv s}}^H_{a_1,a_2,1}{\rm diag}(\mv \theta){\tilde{\mv h}}_{a_1,2}\rvert &{\text{if}}\;l=1\\
	\arg \mathop{\max}\limits_{{\mv \theta} \in {\cal W}_I}\lvert\tilde{\mv g}^H_{a_L,J+k}{\rm diag}(\mv \theta){\tilde{\mv s}}_{a_{L-1},a_L,2}\rvert \!\!&{\text{if}}\;l=L\\
	\arg \mathop{\max}\limits_{{\mv \theta} \in {\cal W}_I}\lvert{\tilde{\mv s}}^H_{a_l,a_{l+1},1}{\rm diag}(\mv \theta){\tilde{\mv s}}_{a_{l-1},a_l,2}\rvert \!\!&{\text{otherwise}},\label{pbf}
\end{cases}
\end{align}\endgroup
where we have defined ${\mv w}_B(j)$ as the optimal active beamforming vector for the BS to transmit the beam to IRS $j$, and ${\mv \theta}(i,j,r)$ as the optimal passive beamforming vector for IRS $j$ to reflect the beam from its previous node $i$ to next node $r$ in its corresponding reflection path.

Let $\bar A_l$ denote the value of $A_l$ in (\ref{Al}) with the optimal passive beamforming design in (\ref{pbf}). Then, by substituting (\ref{abf}) and (\ref{pbf}) into (\ref{e2eCh1}), the maximum BS-EU $k$ channel power gain in any reflection path $\Omega$ is obtained as
\begin{equation}\label{maxCH1}
H_{0,J+k}(\Omega)=\kappa_{0,J+k}(\Omega)\Big(\prod\nolimits_{l=1}^{L}\lvert{\bar A}_l\rvert^2\Big)\lvert{\tilde{\mv h}}^H_{a_1,1}{\mv w}_B(a_1)\rvert^2.
\end{equation}

To simplify the BS and IRS beamforming designs, we consider herein that a set of node-disjoint reflection paths are selected for each EU. Specifically, denote by $\Gamma_k \subseteq \tilde\Gamma_k$ the set of reflection paths selected for EU $k$. Then, for any two paths in $\Gamma_k, k \in {\cal K}$ (e.g., $\Omega_1, \Omega_2 \in \tilde\Gamma_k$), it must hold that $\Omega_1 \cap \Omega_2 = \emptyset$. Furthermore, we consider that the reflection paths associated with different EUs have no common IRS as well, i.e., $\Gamma_k \cap \Gamma_{k'}=\emptyset, k, k' \in {\cal K}, k \ne k'$. It then follows that any two reflection paths in the network (for the same EU or different EUs) are separated, and thus the IRS passive beamforming design in different paths can be decoupled. Let $\Omega_k^{(q)}=\{a^{(q)}_{k,1},a^{(q)}_{k,2},\cdots,a^{(q)}_{k,L_k^{(q)}}\}$ denote the $q$-th reflection path in $\Gamma_k$, where $L_k^{(q)} \ge 1$ and $a^{(q)}_{k,l} \in \cal J$ denote the number of IRSs and the index of the $l$-th IRS in this path, respectively. To ensure the above path separation, we need to have
\begin{align}
&\Omega_k^{(q)} \cap \Omega_k^{(q')} =\emptyset, \forall q,q' \in {\cal Q}_k, q \ne q', k \in {\cal K}, 	\nonumber\\
&\Omega_k^{(q)} \cap \Omega_{k'}^{(q')} =\emptyset, \forall q \in {\cal Q}_k, q' \in {\cal Q}_{k'},	k,k' \in {\cal K}, k \ne k', \label{separate}
\end{align}
where ${\cal Q}_k \triangleq \{1,2,\cdots,\lvert \Gamma_k \rvert\}$. It can be easily shown that for any given BS beamforming vector ${\mv w}_B$, the IRS passive beamforming vectors that maximize $\lvert h_{0,J+k}(\Omega_k^{(q)}) \rvert^2$ can be obtained similarly as (\ref{pbf}), by replacing $a_l$ and $L$ therein with $a^{(q)}_{k,l}$ and $L_k^{(q)}$, respectively.

On the other hand, by exploiting the high angular resolution of the BS with a large $N_B$ (e.g., in massive MIMO), its active beamforming design can be decoupled among different paths as well. Specifically, let ${\cal N}_0=\{j|e_{0,j} \in E_L\}$ denote the set of all first reflecting IRSs in the system (i.e., the IRSs having LoS links with the BS). Evidently, it must hold that $a^{(q)}_{k,1} \in {\cal N}_0, \forall q \in {\cal Q}_k, k \in {\cal K}$. If $N_B$ is sufficiently large and the IRSs in ${\cal N}_0$ are sufficiently separated in the angular domain, we have\cite{ngo2014aspects}
\begingroup\makeatletter\def\f@size{9.8}\check@mathfonts
\def\maketag@@@#1{\hbox{\m@th\normalsize\normalfont#1}}%
\begin{equation}\label{cond}
\frac{1}{N_B}{\lvert\tilde{\mv h}}^H_{j,1}{\mv w}_B(j)\rvert^2 \!\!\approx\!\! 1,\;\frac{1}{N_B}{\lvert\tilde{\mv h}}^H_{j,1}{\mv w}_B(i)\rvert^2 \!\approx\! 0, \forall i,j \in {\cal N}_0, i \ne j,
\end{equation}\endgroup
which implies that the sidelobe effect of ${\mv w}_B(i), i \in {\cal N}_0$ on all first reflecting IRSs (except IRS $i$) can be negligible. Hence, similar to \cite{mei2022split}, we consider that the BS splits its beamforming vector ${\mv w}_B$ into multiple beams with different powers, each pointing to one of its nearby IRSs, i.e.,
\begin{equation}\label{wb}
{\mv w}_B={\sum\limits_{j \in {\cal N}_0}\sqrt{\alpha_j}{\mv w}_B(j)},
\end{equation}
where we assume that the BS's transmit power is unity, and $\alpha_j$ denotes the transmit power allocated to ${\mv w}_B(j)$, with $\sum\nolimits_{j \in {\cal N}_0} \alpha_j=1$. It follows from (\ref{cond}) that $\frac{1}{N_B}{\lvert\tilde{\mv h}}^H_{j,1}{\mv w}_B\rvert^2 \approx \alpha_j, j \in {\cal N}_0$. As such, the BS beamforming vector ${\mv w}_B$ can be decoupled among different paths via their first reflecting IRSs, $a^{(q)}_{k,1} \in {\cal N}_0, q \in {\cal Q}_k, k \in {\cal K}$.

With the above BS and IRS beamforming designs, the end-to-end channel power gain for the path $\Omega_k^{(q)}$ is given by $\alpha_{a^{(q)}_{k,1}}H_{0,J+k}(\Omega_k^{(q)})$. By applying the passive beam combining technique\cite{mei2022split} at the receiver of each EU $k$, the signals over all paths in $\Gamma_k$ can be coherently combined, and its maximum received power is given by
\begin{align}
\tilde E_k &=\Bigg(\sum\limits_{q \in {\cal Q}_k}\sqrt{\alpha_{a^{(q)}_{k,1}}H_{0,J+k}(\Omega_k^{(q)})}\Bigg)^2 \nonumber\\
&\le \rho_k\sum\limits_{q \in {\cal Q}_k}H_{0,J+k}(\Omega_k^{(q)})\triangleq \tilde E_{k,\max}, k \in {\cal K},\label{Emax}
\end{align}
where $\rho_k=\sum\nolimits_{q \in {\cal Q}_k} {\alpha_{a^{(q)}_{k,1}}}$ denotes the total transmit power allocated to EU $k$ (or its associated first reflecting IRSs); the first equality is due to the coherent passive beam combining at the EU; and the inequality is due to the Cauchy-Schwarz inequality by setting $\alpha_{a^{(q)}_{k,1}}=\frac{\rho_k H_{0,J+k}(\Omega_k^{(q)}))}{\sum\nolimits_{q \in {\cal Q}_k}H_{0,J+k}(\Omega_k^{(q)}))}$.

However, the passive beamforming design in (\ref{pbf}) may result in inter-beam interference via the LoS links between any two reflection paths due to the side-lobe of each passive beam in practice owing to its finite spatial resolution. Such ``leaked'' energy signals are generally uncontrolled and may add to their controlled counterparts in a constructive or destructive manner at the EU, thus affecting the designed maximum received signal power given in (\ref{Emax}), $\tilde E_{k,\max}$. To minimize such uncontrolled effects, we propose to apply randomly generated common phase shift at IRS, as elaborated next.\vspace{-9pt}

\subsection{Random Common Phase Shift at Each IRS}\label{rcps}
Specifically, we append the optimized passive beamforming vector of some IRSs with a random common phase shift that varies with time. By this means, all uncontrolled energy signals will randomly add to their controlled counterpart to be averaged out over time, without compromising the strength of the latter at each EU. Let $\omega_j$ denote the common phase shift appended to IRS $j$, which is uniformly distributed within $[0,2\pi)$ and randomly generated at each time. Then, the overall common phase shift appended to the path $\Omega_k^{(q)}$ is given by
\begin{equation}\label{common1}
\omega(\Omega_k^{(q)})=\sum\limits_{l=1}^{L_k^{(q)}}\omega_{a^{(q)}_{k,l}},
\end{equation}
which does not affect the channel power gain $H_{0,J+k}(\Omega_k^{(q)})$. However, to ensure the coherent combining of the received signals over all paths in $\Gamma_k$ at EU $k$, it should be satisfied that the overall common phase shifts appended to all different paths in $\Gamma_k$ are equal, i.e.,
\begin{equation}\label{common2}
\omega(\Omega_k^{(q)})=\omega(\Omega_k^{(q')}), \forall q \ne q', q, q' \in {\cal Q}_k, k \in {\cal K}. 
\end{equation}

For the equalities in (\ref{common2}) to hold, as each EU $k, k \in \cal K$ is associated with $\lvert \Gamma_k \rvert$ reflection paths, there should exist $\lvert \Gamma_k \rvert-1$ equations w.r.t. $\sum\nolimits_{q \in {\cal Q}_k}L_k^{(q)}$ common phase shifts (or variables) that should be met. Thus, for all $K$ EUs, there should exist $\sum\nolimits_{k=1}^{K}{\lvert \Gamma_k \rvert}-K$ equations in (\ref{common2}), w.r.t. $\sum\nolimits_{k=1}^{K}\sum\nolimits_{q \in {\cal Q}_k}L_k^{(q)}$ common phase shifts. Since $L_k^{(q)} \ge 1$, we have
\[\sum\limits_{k=1}^{K}\sum\limits_{q \in {\cal Q}_k}L_k^{(q)} \ge \sum\limits_{k=1}^{K}{\lvert \Gamma_k \rvert} > \sum\limits_{k=1}^{K}{\lvert \Gamma_k \rvert}-K,\]
which indicates that the total number of common phase shifts (or IRSs involved) must be larger than that of the equations. As such, there exist infinitely many solutions to (\ref{common2}). In particular, we can randomly set the common phase shifts of at least $K$ IRSs and then determine those of other IRSs accordingly. By this means, for each EU $k, k \in \cal K$, the overall phase of any interference path from the BS to it (i.e., the path in $\tilde\Gamma_k \backslash \Gamma_k$) can be assumed to be independent to that of any signal path in $\Gamma_k$ over time. Thus, it can be shown that the average received signal power by EU $k$ in the long term is
 \begin{equation}\label{actual}
 	\tilde E_{k,{\text{av}}} = \tilde E_{k,\max} + \sum\limits_{\Omega \in \tilde\Gamma_k \backslash \Gamma_k} \lvert h_{0,J+k}(\Omega) \rvert^2, k \in {\cal K},
 \end{equation}
which is larger than $\tilde E_{k,\max}$ in (\ref{Emax}). However, due to the lack of CPB gains over the randomly scattered interference paths in $\tilde\Gamma_k \backslash \Gamma_k$, their strength is generally much lower than that of the signal paths in $\Gamma_k$ and thus can be ignored\cite{mei2022mbmh}. Hence, we assume in this paper $\tilde E_{k,{\text{av}}} \approx \tilde E_{k,\max}$.

\begin{figure}[!t]
\centering
\includegraphics[width=3in]{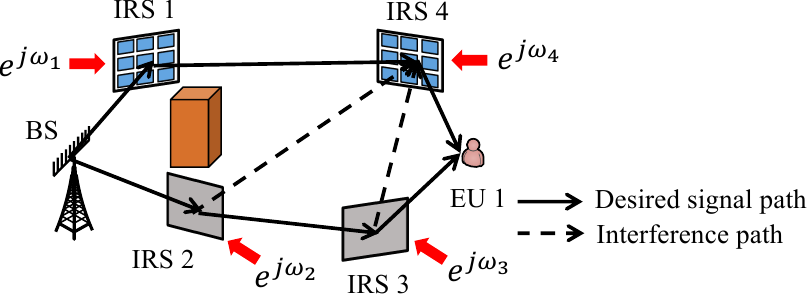}
\DeclareGraphicsExtensions.
\caption{An illustration of the random common phase shift.}\label{NumExp}
\vspace{-15pt}
\end{figure}
As an illustrative example, we show in Fig.\,\ref{NumExp} two reflection paths that are selected for EU 1, denoted as $\Omega_1^{(1)}=\{1,4\}$ and $\Omega_1^{(2)}=\{2,3\}$, respectively. Note that both IRSs 2 and 3 in the path $\Omega_1^{(2)}$ can randomly scatter interference to path $\Omega_1^{(1)}$ via the LoS link from them to IRS 4, resulting in two interference paths (in addition to the desired signal paths $\Omega_1^{(1)}$ and $\Omega_1^{(2)}$), denoted as $S_1=\{2,4\}$ and $S_2=\{2,3,4\}$. Based on (\ref{common2}), we need to satisfy $\omega_1+\omega_4=\omega_2+\omega_3$. Thus, we can randomly generate $\omega_2$, $\omega_3$, and $\omega_4$ each time and set $\omega_1=\omega_2+\omega_3-\omega_4$. Let $\psi(\Omega)$ denote the phase shift of any path $\Omega$ before appending the common phase shifts to its constituent IRSs, with $\psi(\Omega_1^{(1)})=\psi(\Omega_1^{(2)})$. Then, after appending them, the phase shifts of the signal paths $\Omega_1^{(1)}$ and $\Omega_1^{(2)}$ become $\omega_1+\omega_4+\psi(\Omega_1^{(1)})=\omega_2+\omega_3+\psi(\Omega_1^{(2)})$, while those of the interference paths $S_1$ and $S_2$ are given by $\omega_2+\omega_4+\psi(S_1)$ and $\omega_2+\omega_3+\omega_4+\psi(S_2)$, respectively. It follows that the phase difference between $\Omega_1^{(1)}$ (or $\Omega_1^{(2)}$) and $S_1$ is given by $\omega_3-\omega_4+\psi(\Omega_1^{(2)})-\psi(S_1)$. Since $\psi(\Omega_1^{(2)})$ and $\psi(S_1)$ are fixed (as determined by the optimized phase shifts of the IRSs) while $\omega_3$ and $\omega_4$ are independently generated, the received signals over these two interference paths will be randomly added at the receiver of EU 1. Similarly, it can be shown that the paths $\Omega_1^{(1)}$ (or $\Omega_1^{(2)}$) and $S_2$, as well as the paths $S_1$ and $S_2$ are randomly added as well. In contrast, only the signals in the desired paths $\Omega_1^{(1)}$ and $\Omega_1^{(1)}$ are coherently added. \vspace{-9pt}

\subsection{Beam Routing Problem Formulation}
Given the maximum received power in (\ref{Emax}), we aim to jointly optimize the reflection paths for all EUs ($\Gamma_k, k \in \cal K$) and the BS power allocation among all EUs ($\rho_k, k \in {\cal K}$) to maximize the minimum received power among all EUs, i,e., $\mathop {\min}\limits_{k \in \cal K} \tilde E_{k,\max}$. The corresponding optimization problem is given by
\begin{align}\label{op1}
{\text{(P1)}} &\mathop {\max}\limits_{\{\Gamma_k,\rho_k\}_{k \in {\cal K}}}\;\mathop {\min}\limits_{k \in \cal K}\;\tilde E_{k,\max} \nonumber\\
\text{s.t.}\;\;&e(a^{(q)}_{k,l},a^{(q)}_{k,l+1}) \in E_L, \forall 0 \le l \le L_k^{(q)}, q \in {\cal Q}_k, k \in {\cal K},\nonumber\\
&\Omega_k^{(q)} \cap \Omega_k^{(q')} =\emptyset, \forall q,q' \in {\cal Q}_k, q \ne q', k \in {\cal K}, 	\nonumber\\
&\Omega_k^{(q)} \cap \Omega_{k'}^{(q')} =\emptyset, \forall q \in {\cal Q}_k, q' \in {\cal Q}_{k'},	k,k' \in {\cal K}, k \ne k',\nonumber\\
& \sum\limits_{k \in \cal K} \rho_k =1, \rho_k \ge 0, k \in {\cal K}.
\end{align}

Note that problem (P1) is more general than the beam routing problems studied in our prior works \cite{mei2022mbmh} and \cite{mei2022split}, which only consider single-path beam routing and single EU, respectively. In practice, (P1) can be solved at the BS based on the channel information feedback from distributed IRSs via their offline beam training\cite{mei2021distributed}. The solution to (P1) will be presented later in Section \ref{pf}.

Note that the baseline beam routing design in (P1) may result in suboptimal WPT performance, as each IRS only serves at most a single EU. In practice, each IRS can generally serve multiple EUs by modifying the baseline scheme. For example, the IRS passive beamforming and beam routing can be dynamically varied to serve different EUs over orthogonal time slots, such that each EU can be served by more IRSs to reap higher CPB and LoS path diversity gains. Alternatively, each IRS can split its reflecting elements into multiple subsurfaces to reflect the incident signals to multiple nearby IRSs/EUs in different directions, thereby creating more signal paths to serve multiple EUs at the same time. Motivated by the above, we propose two new beam routing schemes for WPT in the next section, where each IRS is allowed to serve multiple EUs over different time and different directions, respectively.\vspace{-6pt}

\section{Beam Routing with Each IRS Serving Multiple EUs}\label{proposed}
In this section, we present two new beam routing schemes, namely, dynamic and subsurface-based beam routing schemes. For both schemes, we derive their respective optimal active and passive beamforming at the BS and each involved IRS for each reflection path, based on which the average received signal power at each EU is derived. Then, we show that the dynamic beam routing in general outperforms the subsurface-based beam routing.\vspace{-9pt}

\subsection{Dynamic Beam Routing}
\begin{figure}[!t]
\centering
\includegraphics[width=3.4in]{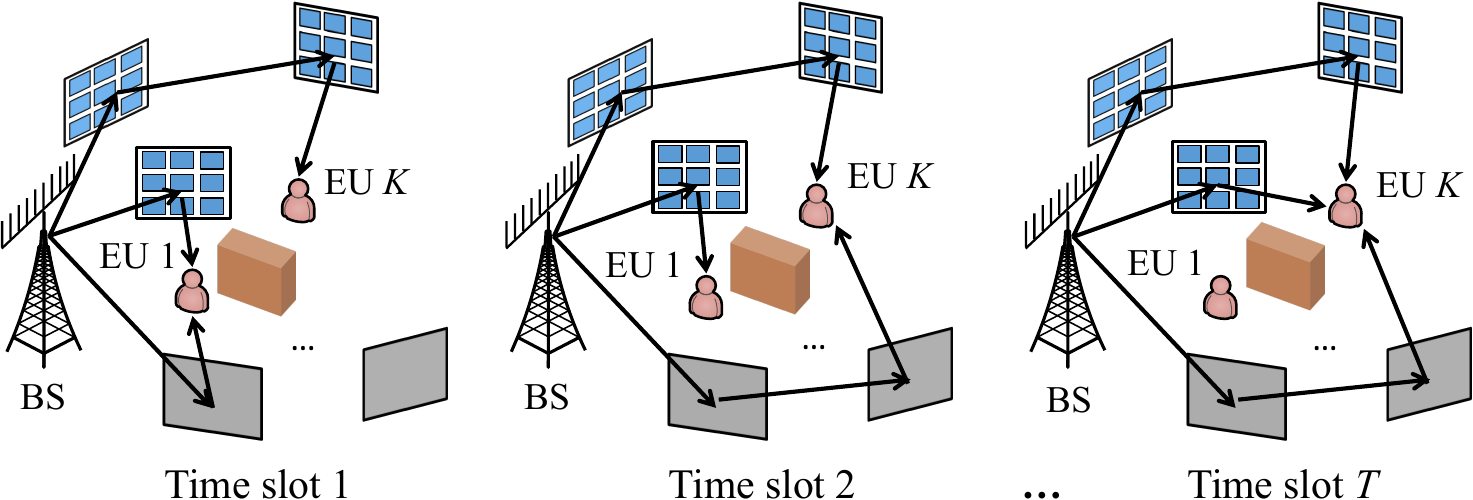}
\DeclareGraphicsExtensions.
\caption{Illustration of the dynamic beam routing scheme (with only EUs 1 and $K$ shown).}\label{dynamic}
\vspace{-12pt}
\end{figure}
In this scheme, we exploit the time selectivity of IRS passive beamforming and the resulted time-varying beam-routing paths for improving the multi-EU WPT performance. Specifically, as illustrated in Fig.\,\ref{dynamic}, we consider that the total WPT duration is divided into $T$ time slots, over which the reflection paths can be varied to serve different sets of EUs. For each time slot $t, t \in {\cal T}\triangleq \{1,2,\cdots,T\}$, denote $\tau_t$ and $\Gamma_{k,t}$ as its duration and the set of reflection paths selected for EU $k, k \in {\cal K}$, respectively, with $\sum\nolimits_{t=1}^T{\tau_t}=1$ (assuming the total duration is equal to one). Note that if $\Gamma_{k,t}=\emptyset$, then EU $k$ is not assigned for WPT in time slot $t$. Similar to the beam routing design in Section \ref{conv}, we assume that the paths selected for each EU and for different EUs (i.e., the paths in the set $\bigcup_{k \in \cal K}\Gamma_{k,t}$) are both node-disjoint in each time slot $t$, as shown in Fig.\,\ref{dynamic}. Hence, the reflection paths in each time slot correspond to a feasible solution to (P1).

Given the reflection paths $\Gamma_{k,t}, k \in {\cal K}$ in time slot $t$, we can optimize the BS and IRS beamforming similarly as in Section \ref{conv}. Let $\rho_{k,t}$ denote the BS's transmit power allocated to EU $k$ in time slot $t$. Then, based on (\ref{Emax}), the total power received by EU $k$ in time slot $t$ is given by
\begin{equation}\label{EkMax}
E_{k,t}=\rho_{k,t}\Bigg(\sum\limits_{\Omega \in \Gamma_{k,t}}H_{0,J+k}(\Omega)\Bigg), k \in {\cal K}, t \in {\cal T},
\end{equation}
and the average received power over all $T$ time slots is given by
\begin{equation}\label{EkSW}
E_{k,{\text{D}}}=\sum\limits_{t \in {\cal T}}{\tau_t\rho_{k,t}\Bigg(\sum\limits_{\Omega \in \Gamma_{k,t}}H_{0,J+k}(\Omega)\Bigg)}.
\end{equation}

It follows that for any given reflection path selection over the $T$ time slots, i.e., $\Gamma_{k,t}, k \in {\cal K}, t \in {\cal T}$, we should determine the optimal time allocation $\tau_t, t \in \cal T$ for all time slots and the optimal power allocation $\rho_{k,t}, k \in {\cal K}$ in each time slot $k$. Hence, the optimal performance of the dynamic beam routing can be achieved by accounting for all possible combinations of reflection path sets for the $K$ EUs (i.e., all feasible solutions to (P1)) and determining the optimal time allocations among them. Accordingly, the number of time slots, $T$, should be set as that of feasible solutions to (P1). For example, one possible combination is such that only one EU (e.g., EU 1) is assigned for WPT in time slot $t$, i.e., $\Gamma_{k,t}=\emptyset, k \ne 1, k\in {\cal K}$. Based on the above, we can redefine $\Gamma_{k,t}, k \in \cal K$ as the $t$-th feasible solution to (P1). Note that if $\tau_t=0$, then the reflection paths in this solution will not be selected to serve any EU.\vspace{-6pt}

\subsection{Subsurface-Based Beam Routing}
\begin{figure}[!t]
\centering
\includegraphics[width=2.4in]{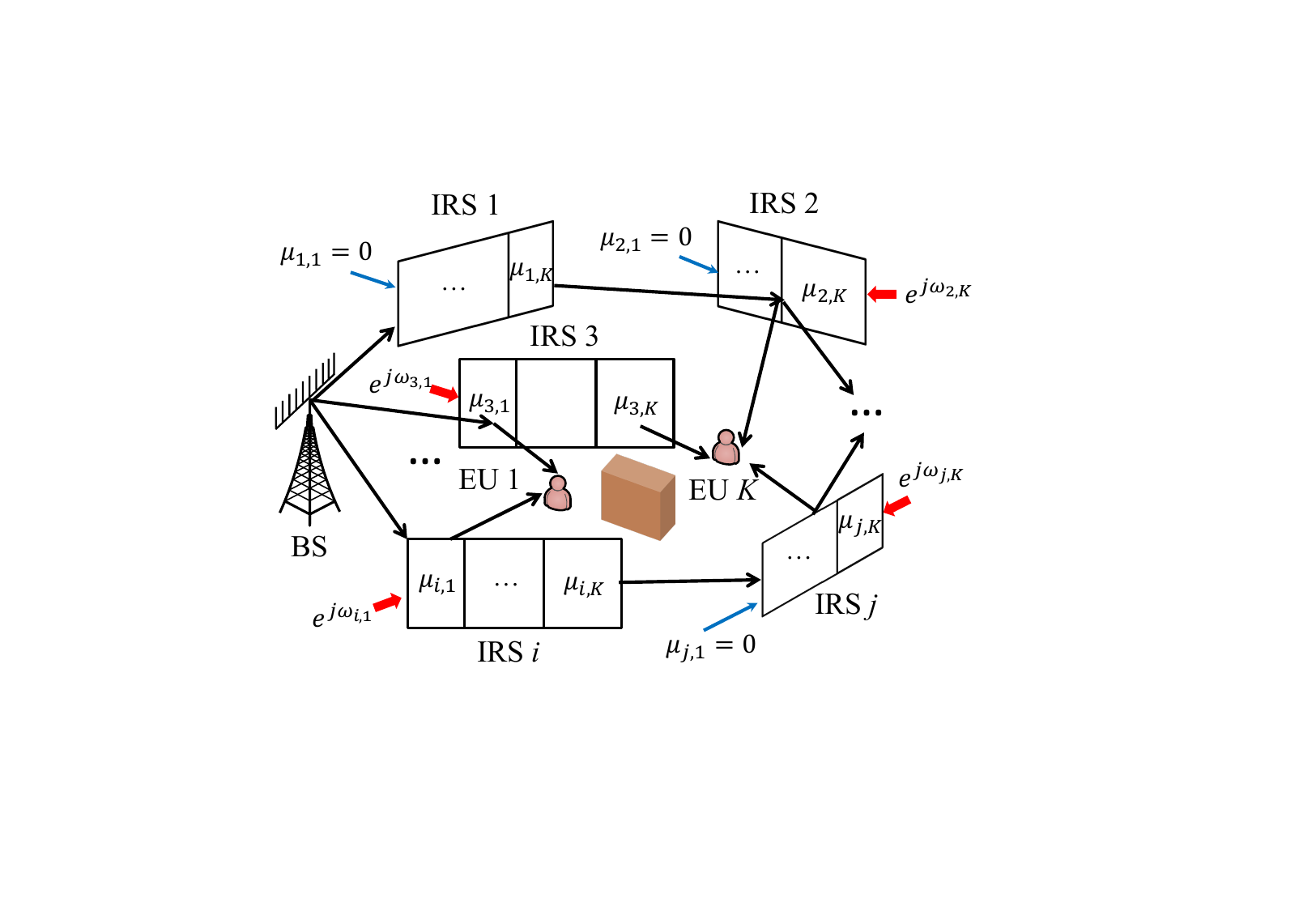}
\DeclareGraphicsExtensions.
\caption{Illustration of the subsurface-based beam routing scheme (with only EUs 1 and $K$ shown).}\label{subsurface}
\vspace{-9pt}
\end{figure}
In this scheme, we adopt the subsurface-based beamforming at each IRS to serve multiple EUs simultaneously. Specifically, as illustrated in Fig.\,\ref{subsurface}, we split the reflecting elements of each IRS $j, j \in \cal J$ into $K$ subsurfaces, each associated with one of the $K$ EUs. By this means, more signal paths can be created, such that each IRS is able to serve multiple EUs at the same time (e.g., IRSs 3 and $i$ in Fig.\,\ref{subsurface}). For convenience, we assume that each IRS is split along its horizontal direction (i.e., the $x$-axis direction in its local coordinate system). Let $\mu_{j,k} \in [0,1]$ denote the fraction of the horizontal elements of IRS $j, j \in \cal J$ allocated to EU $k, k \in \cal K$, with $\sum\nolimits_{k \in \cal K}\mu_{j,k}=1$. For convenience, we label the $k$-th subsurface of IRS $j$ as IRS $j_k$, with $\mu_{j,k}M$ reflecting elements. Note that it is possible that $\mu_{j,k}=0$ and $\mu_{j,k}=1$ if none and all of the reflecting elements of IRS $j$ are allocated to EU $k$, respectively. For example, in Fig.\,\ref{subsurface}, IRSs 1, 2, and $j$ are not used to serve EU 1, and we have $\mu_{1,1}=\mu_{2,1}=\mu_{j,1}=0$. Then, the subsurface-based channels can be modeled as follows. If $e_{0,j} \in E_L$, the BS-IRS $j_k$ channel is expressed as
$
{\mv H}_{0,j_k} = \frac{\sqrt \beta}{d_{0,j}}e^{-\frac{j2\pi d_{0,j}}{\lambda}}{\tilde{\mv h}}^{(k)}_{j,2}{\tilde{\mv h}}^H_{j,1}, \;k \in {\cal K}, j \in {\cal J},
$
where ${\tilde{\mv h}}^{(k)}_{j,2}={\tilde{\mv h}}_{j,2}(1+M\sum\nolimits_{s=0}^{k-1}\mu_{j,s}:M\sum\nolimits_{s=0}^{k}\mu_{j,s})$. Here, we set $\mu_{j,0}=0, j \in \cal J$, and ${\mv h}(a:b)$ extracts the $a$-th-to-$b$-th elements of the vector $\mv h$. Similarly, if $e_{i,j} \in E_L, i,j \in \cal J$, the IRS $i_k$-IRS $j_l$ channel is given by
$
{\mv S}_{i_k,j_l} = \frac{\sqrt \beta}{d_{i,j}}e^{-\frac{j2\pi d_{i,j}}{\lambda}}{\tilde{\mv s}}^{(l)}_{i,j,2}{\tilde{\mv s}}^{(k)H}_{i,j,1}, \;i, j \in {\cal J}, i \ne j, k,l \in {\cal K}.
$
where ${\tilde{\mv s}}^{(l)}_{i,j,2}={\tilde{\mv s}}_{i,j,2}(1+M\sum\nolimits_{s=0}^{l-1}\mu_{j,s}:M\sum\nolimits_{s=0}^{l}\mu_{j,s})$ and ${\tilde{\mv s}}^{(k)}_{i,j,1}={\tilde{\mv s}}_{i,j,1}(1+M\sum\nolimits_{s=0}^{k-1}\mu_{i,s}:M\sum\nolimits_{s=0}^{k}\mu_{i,s})$.
Finally, if $e_{j,J+k} \in E_L$, the IRS $j_l$-EU $k$ channel is expressed as
$
{\mv g}^H_{j_l,J+k} \!=\! \frac{\sqrt \beta}{d_{j,J+k}}e^{-\frac{j2\pi d_{{j,J+k}}}{\lambda}}{\tilde{\mv g}}^{(l)H}_{j,J+k}, \;j \in {\cal J}, k \in {\cal K},
$
where ${\tilde{\mv g}}^{(l)H}_{j,J+k}={\tilde{\mv g}}^H_{j,J+k}(1+M\sum\nolimits_{s=0}^{l-1}\mu_{j,s}:M\sum\nolimits_{s=0}^{l}\mu_{j,s})$.

It is worth noting that given the element splitting ratios $\mu_{j,k}$'s, the reflection paths for all EUs can be decoupled. In particular, for each EU $k, k \in \cal K$, we can consider an equivalent multi-IRS system comprising all subsurfaces in ${\cal J}_k \triangleq \{j_k\}_{j \in \cal J}$. Similarly as in Section \ref{conv}, we select a set of node-disjoint reflection paths from the subsurfaces in ${\cal J}_k$ to serve each EU $k, k \in \cal K$, denoted as $\Upsilon_k$. Then, let the $r$-th reflection path in $\Upsilon_k$ be expressed as $B_k^{(r)}=\{b^{(r)}_{k,1},b^{(r)}_{k,2},\cdots,b^{(r)}_{k,S_k^{(r)}}\}$, where $S_k^{(r)} \ge 1$ and $b^{(r)}_{k,s} \in {\cal J}_k$ denote the number of subsurfaces and the index of the $s$-th subsurface in this path, respectively, with $b^{(r)}_{k,1} \in \{j_k\}_{j \in {\cal N}_0}$. Next, for any given BS active beamforming ${\mv w}_B$, the subsurface beamforming that maximizes the end-to-end channel power gain over the path $B_k^{(r)}$ can be obtained similarly as (\ref{pbf}), by replacing $a_l$ and $L$ therein with $b^{(r)}_{k,l}$ and $S_k^{(r)}$, respectively. With the BS's active beamforming in (\ref{wb}) and the coherent combining at each EU $k$, its received signal power can be expressed as
\begin{equation}\label{EkSP}
E_{k,{\text{S}}}=\Bigg(\sum\limits_{r \in {\cal R}_k}\sqrt{\alpha_{b^{(r)}_{k,1}}H_{0,J+k}(B_k^{(r)})}\Bigg)^2, k \in {\cal K},
\end{equation}
where ${\cal R}_k=\{1,2,\cdots,\lvert \Upsilon_k \rvert\}$ and $\alpha_{b^{(r)}_{k,1}}$ denotes the BS's transmit power allocated to the first reflecting subsurface $b^{(r)}_{k,1}$. It should be noted that for each first reflecting subsurface $j_k, j \in {\cal N}_0$, as it belongs to IRS $j$, we have $\alpha_{j_k}=\alpha_j, k \in {\cal K}$. However, different from (\ref{Emax}), the Cauchy-Schwarz inequality cannot be used for (\ref{EkSP}) to decouple the power allocation ratios $\alpha_j, j \in {\cal N}_0$ among all EUs as $\rho_k, k \in \cal K$. This is because each first reflecting IRS in ${\cal N}_0$ is shared by multiple EUs at the same time under the subsurface-based beamforming. 

It is also worth mentioning that in the subsurface-based beam routing scheme, the inter-path interference may have a more significant effect on the WPT performance compared to the dynamic beam routing scheme, due to the smaller size of each subsurface, which has a more severe side-lobe effect. Nonetheless, as shown in Fig.\,\ref{subsurface}, we can still append a random common phase shift (denoted as $\omega_{j,k}$) to some subsurfaces $j_k, j \in {\cal J}, k \in {\cal K}$, thereby ensuring the WPT performance in (\ref{EkSP}) in the long term, similar to our discussion in Section \ref{rcps}.\vspace{-9pt}

\subsection{Performance Comparison}
In this subsection, we compare the performance of the previously presented three beam routing schemes, namely, the baseline beam routing with each IRS serving at most one EU, dynamic beam routing, and subsurface-based beam routing. First, it is noted that the latter two schemes can both include the former one as a special case by further constraining $T=1$ and $\mu_{j,k} \in \{0,1\}, \forall j \in {\cal J}, k \in {\cal K}$, respectively. Hence, their performance must be no worse than that of the former one. As such, we only compare the performance of the latter two schemes. For ease of exposition, we consider that the codebook size of passive beamforming at each IRS is infinity, i.e., $\lvert {\cal W}_I \rvert \rightarrow \infty$, so that each IRS reflecting element can adopt any continuous phase shift. In this case, it can be easily shown that ${\bar A}_l=M$ in (\ref{Al}); hence, the maximum channel power gain in (\ref{maxCH1}) can be simplified as
\begin{equation}\label{maxCH2}
H_{0,J+k}(\Omega)=\kappa_{0,J+k}(\Omega)M^{2L}N_B.
\end{equation}
While for a subsurface-based reflection path (e.g., $B_k^{(r)}$ in (\ref{EkSP})), its maximum channel power gain is given by
\begin{equation}\label{maxCH3}
H_{0,J+k}(B_k^{(r)})=\Bigg(\prod\limits_{j \in B_k^{(r)}}{\mu^2_{j,k}}\Bigg)N_BM^{2S_k^{(r)}}\kappa_{0,J+k}(B_k^{(r)}),
\end{equation}
which experiences a multiplicative loss in CPB gain due to the IRS element splitting.

\begin{figure}[!t]
\centering
\includegraphics[width=2.6in]{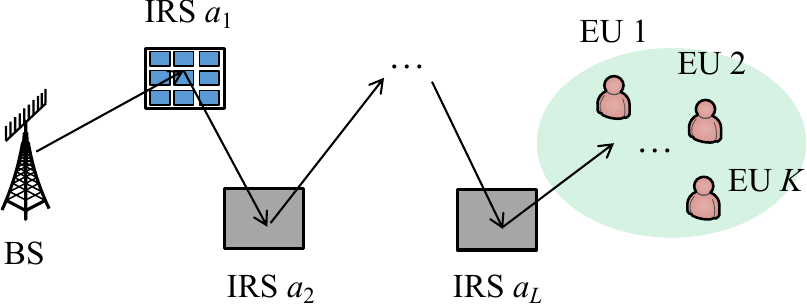}
\DeclareGraphicsExtensions.
\caption{A simplified scenario to show the superiority of the dynamic beam routing scheme.}\label{SimpleExample}
\vspace{-12pt}
\end{figure}
Next, we show that the dynamic beam routing can yield a performance no worse than that of the subsurface-based beam routing. This can be first intuitively explained by considering a simplified scenario, where the BS serves all $K$ EUs via the reflection path $\Omega=\{a_1,a_2,\cdots,a_L\}$ in (\ref{e2eCh1}), as shown in Fig.\,\ref{SimpleExample}. As the $K$ EUs can only achieve LoS links with IRS $a_L$ and the BS can only achieve an LoS link with IRS $a_1$, in the subsurface-based beam routing scheme, we only need to determine the element splitting ratios, $\mu_{a_L,k}, k \in {\cal K}$, with $\sum\nolimits_{k \in \cal K}{\mu_{a_L,k}}=1$. While in the dynamic beam routing, the path $\Omega$ can be used to serve each of the $K$ EUs exclusively over $K$ orthogonal time slots. Hence, we only need to determine the time allocation ratios for the $K$ EUs, i.e., $\tau_k, k \in \cal K$, with $\sum\nolimits_{k \in \cal K}{\tau_k}=1$. Then, based on (\ref{maxCH2}) and (\ref{maxCH3}), it can be verified that under the subsurface-based and dynamic beam routing schemes, the received signal power by EU $k$ is given by $E_{k,{\text{S}}}=\mu^2_{a_L,k}\kappa_{0,J+k}(\Omega)M^{2L}N_B$ and $E_{k,{\text{D}}}=\tau_{k}\kappa_{0,J+k}(\Omega)M^{2L}N_B$, respectively. It is evident that by setting $\tau_k=\mu_{a_L,k}$, we have $E_{k,{\text{D}}} \ge E_{k,{\text{S}}}$ due to $\mu_{a_L,k} \ge \mu^2_{a_L,k}$. This implies that the dynamic beam routing scheme can outperform the subsurface-based one, as the former can achieve the full CPB gain for each EU, which overwhelms the loss due to the reduced WPT time. 

For the general case, the above result still holds, as presented in the following proposition.
\begin{proposition}\label{prop}
With continuous IRS phase shifts, it holds that $E_{k,\text{D}} \ge E_{k,\text{S}}, \forall k \in \cal K$, i.e., the dynamic beam routing can achieve a minimum received power among all EUs no smaller than that by the subsurface-based beam routing.
\end{proposition}
\begin{IEEEproof}
We need to show that in the subsurface-based beam routing scheme, for any given reflection paths for all EUs ($\Upsilon_k, k \in \cal K$), element splitting ratios ($\mu_{j,k}, j \in {\cal J}, k \in {\cal K}$), and BS power allocations among all first reflecting IRSs ($\alpha_j, j \in {\cal N}_0$), we can construct a dynamic beam routing solution accordingly, such that $E_{k,\text{D}} \ge E_{k,\text{S}}, \forall k \in {\cal K}$ is met. Specifically, we consider $T=K$ time slots in the dynamic beam routing scheme, each for WPT to one of the $K$ EUs, respectively. Without loss of generality, we assume that EU $k$ is assigned for WPT in time slot $k$. Let $\Pi(B_k^{(r)})$ denote the path that replaces each subsurface in $B_k^{(r)}$ with its associated IRS. Then, in time slot $k$, we consider that the reflection paths $\Pi(B_k^{(r)}), r \in {\cal R}_k$ are selected to serve EU $k$. As a result, based on (\ref{maxCH2}), its received power is given by
\begin{equation}\label{E1}
E_{k,\text{D}}=\tau_k\sum\limits_{r \in {\cal R}_k}N_BM^{2S_k^{(r)}}\kappa_{0,J+k}(B_k^{(r)}), k \in {\cal K}.
\end{equation}
On the other hand, its received power under the subsurface-based beam routing scheme is given by
\begingroup\makeatletter\def\f@size{9.2}\check@mathfonts
\def\maketag@@@#1{\hbox{\m@th\normalsize\normalfont#1}}%
\begin{equation}\label{E2}
E_{k,\text{S}}=\Bigg(\sum\limits_{r \in {\cal R}_k}\sqrt{\Bigg(\prod\limits_{j \in B_k^{(r)}}{\mu^2_{j,k}}\Bigg)\alpha_{b^{(r)}_{k,1}}N_BM^{2S_k^{(r)}}\kappa_{0,J+k}(B_k^{(r)})}\Bigg)^2.
\end{equation}\endgroup
By comparing (\ref{E2}) with (\ref{E1}), it is observed that with the IRS element splitting, each EU $k$ is served by degraded active beamforming and CPB gains over the path $B_k^{(r)}$ (by a factor of $\alpha_{b^{(r)}_{k,1}}$ and $\prod\nolimits_{j \in B_k^{(r)}}{\mu^2_{j,k}}$, respectively), while its time assigned for WPT is enlarged (by a factor of $1/\tau_k$).

By applying the Cauchy-Schwarz inequality to (\ref{E2}), we have
\begingroup\makeatletter\def\f@size{9}\check@mathfonts
\def\maketag@@@#1{\hbox{\m@th\normalsize\normalfont#1}}%
\begin{equation}\label{E3}
E_{k,\text{S}} \le \Bigg(\underbrace{\sum\limits_{r \in {\cal R}_k} \alpha_{b^{(r)}_{k,1}}\prod\limits_{j \in B_k^{(r)}}{\mu^2_{j,k}}}_{\triangleq \epsilon_k}\Bigg)\Bigg(\sum\limits_{r \in {\cal R}_k}N_BM^{2S_k^{(r)}}\kappa_{0,J+k}(B_k^{(r)})\Bigg).
\end{equation}\endgroup

Next, we show that $\sum\nolimits_{k \in {\cal K}}\epsilon_k \le 1$. To this end, note that 
\[\sum\limits_{k \in {\cal K}}\epsilon_k \le \sum\limits_{k \in {\cal K}}\sum\limits_{r \in {\cal R}_k} \alpha_{b^{(r)}_{k,1}}{\mu_{b^{(r)}_{k,1},k}}.\]
Here, we can equivalently view $\alpha_j\mu_{j,k}, j \in {\cal N}_0$ as the BS's power allocated to the $k$-th subsurface of IRS $j$ in the subsurface-based beam routing scheme, with \[\sum\limits_{j \in {\cal N}_0}\sum\limits_{k \in {\cal K}}\alpha_j\mu_{j,k}=\sum\limits_{j \in {\cal N}_0}\alpha_j\Bigg(\sum\limits_{k \in {\cal K}}\mu_{j,k}\Bigg)=\sum\limits_{j \in {\cal N}_0} \alpha_j=1.\]
As a result, we have
\[\sum\limits_{k \in {\cal K}}\epsilon_k \le \sum\limits_{k \in {\cal K}}\sum\limits_{r \in {\cal R}_k} \alpha_{b^{(r)}_{k,1}}{\mu_{b^{(r)}_{k,1},k}} \le \sum\limits_{k \in {\cal K}}\sum\limits_{r \in {\cal N}_0} \alpha_r{\mu_{r,k}}=1,\]
where the second inequality is due to the fact that the set of first reflecting subsurfaces associated with EU $k$ must be the subset of all available first reflecting subsurfaces, i.e., ${\cal R}_k \subseteq {\cal N}_0$.

As $\sum\nolimits_{k \in {\cal K}}\epsilon_k \le 1$, we can set $\tau_k = \epsilon_k$ in (\ref{E1}), under which $E_{k,\text{D}} \ge E_{k,\text{S}}$ can be achieved based on (\ref{E3}). This thus completes the proof.
\end{IEEEproof}

Proposition \ref{prop} implies that for the subsurface-based beam routing, its resultant loss of active beamforming and CPB gains in each reflection path exceeds the gain due to the prolonged time for WPT, which thus results in worse performance than dynamic beam routing. This is because for the end-to-end received signal power at each EU, the loss of WPT time in the dynamic beam routing is inversely linear w.r.t. the number of users, while the loss of the CPB gain is quadratic w.r.t. IRS element splitting ratios. In light of the above result, in the sequel of this paper, we only consider the dynamic beam routing design and solve its corresponding optimization problem. In addition, we refer to the special case of baseline beam routing with each IRS serving at most one EU in Section \ref{conv} as the static beam routing in the sequel for convenience.\vspace{-6pt}

\section{Proposed Solution for Joint Beam Routing and Resource Allocation Optimization}\label{pf}
In this section, we first formulate the design problem for the dynamic beam routing scheme and propose an efficient graph-optimization-based algorithm to solve it, which is then applied to solve (P1) for the static beam routing case.\vspace{-9pt}
\subsection{Problem Formulation}
For the dynamic beam routing scheme, we aim to maximize the minimum received signal power among all EUs by jointly optimizing the time allocations $\tau_t, t \in {\cal T}$ and power allocations $\rho_k, k \in {\cal K}$, i.e.,
\begin{align}\label{op2}
{\text{(P2)}} \mathop {\max}\limits_{\{\tau_t\}, \{\rho_k\}}\;&\mathop {\min}\limits_{k \in \cal K}\;\sum\limits_{t \in {\cal T}}{\tau_t\rho_{k,t}\Bigg(\sum\limits_{\Omega \in \Gamma_{k,t}}H_{0,J+k}(\Omega)\Bigg)} \nonumber\\
\text{s.t.}\;\;&\sum\limits_{t=1}^T{\tau_t}=1, {\tau_t} \ge 0, \forall t \in {\cal T},\nonumber\\
&\sum\limits_{k=1}^K{\rho_{k,t}}=1, \rho_{k,t} \ge 0, \forall k \in {\cal K}, t \in {\cal T}.
\end{align}
It is observed that (P2) is a non-convex optimization problem due to the product of $\tau_t$ and $\rho_{k,t}$ in its objective function. In addition, with increasing $J$ and/or $K$, the number of path combinations (or feasible solutions to (P1)), $T$, becomes larger, thus requiring an efficient path enumeration strategy. Furthermore, as compared to the multi-user/-path beam routing problems studied in prior works (e.g., \cite{mei2022mbmh,mei2022split}), (P2) is more complicated due to the presence of all possible reflection path combinations for all EUs, as well as their intricate coupling with the power and time allocations.\vspace{-6pt} 

\subsection{Proposed Solution to (P2)}\label{solP2}
To solve (P2), it is noted that by introducing an auxiliary variable $\eta_{k,t}=\tau_t\rho_{k,t}, \forall k,t$, (P2) can be equivalently recast as
\begin{align}\label{op4}
{\text{(P2.1)}} \mathop {\max}\limits_{\{\eta_{k,t}\}}\;&\mathop {\min}\limits_{k \in \cal K}\;\sum\limits_{t \in {\cal T}}{\eta_{k,t}\Bigg(\sum\limits_{\Omega \in \Gamma_{k,t}}H_{0,J+k}(\Omega)\Bigg)} \nonumber\\
\text{s.t.}\;\;&\sum\limits_{k=1}^K\sum\limits_{t=1}^T{\eta_{k,t}}=1, \eta_{k,t} \ge 0, \forall k \in {\cal K}, t \in {\cal T}.
\end{align}
Let $\eta^*_{k,t}, k \in {\cal K}, t \in {\cal T}$ denote the optimal solution to (P2.1). Then, the optimal $\tau_t$'s and $\rho_{k,t}$'s for (P2) can be respectively obtained as
\begin{equation}\label{rho}
\tau^*_t = \sum\limits_{k=1}^K{\eta^*_{k,t}}, t \in {\cal T},\quad\rho^*_{k,t} = \frac{\eta^*_{k,t}}{\sum\limits_{k=1}^K{\eta^*_{k,t}}}, k \in {\cal K}, t \in {\cal T}.
\end{equation}
It follows from the above that if all possible reflection path combinations or feasible solutions to (P1), i.e., $\Gamma_{k,t}, k \in {\cal K}, t \in {\cal T}$, are available, it suffices to determine the optimal power allocation for each $\Gamma_{k,t}$, i.e., $\eta_{k,t}$, by solving (P2.1). Next, we show that (P2.1) admits a closed-form solution. To this end, let 
\begin{equation}
H_{k,\max} = \mathop {\max}\limits_{t \in {\cal T}}\sum\limits_{\Omega \in \Gamma_{k,t}}H_{0,J+k}(\Omega),\quad \eta_k = \sum\limits_{t \in {\cal T}}{\eta_{k,t}}, k \in {\cal K}
\end{equation}
with $\sum\limits_{k \in {\cal K}}\eta_k = 1$. Then, we have \[\sum\limits_{t \in {\cal T}}{\eta_{k,t}}\Bigg(\sum\limits_{\Omega \in \Gamma_{k,t}}H_{0,J+k}(\Omega)\Bigg) \le H_{k,\max}\sum\limits_{t \in {\cal T}}{\eta_{k,t}} = H_{k,\max}\eta_k.\]

Based on the above, we can further simplify (P2.1) as
\begin{equation}\label{op5}
{\text{(P2.2)}} \mathop {\max}\limits_{\{\eta_k\}}\mathop {\min}\limits_{k \in \cal K}\;H_{k,\max}\eta_k, \quad\text{s.t.}\;\sum\limits_{k \in {\cal K}}\eta_k = 1.
\end{equation}

Problem (P2.2) can be optimally solved in closed-form. Specifically, it is easy to verify that at the optimality of (P2.2), it must hold that all $H_{k,\max}\eta_k, k \in \cal K$ are identical. Based on this fact, the optimal solution to (P2.2) can be derived as
\begin{equation}\label{opt_eta}
	\eta_k = \frac{H^{-1}_{k,\max}}{\sum\limits_{i \in \cal K}H^{-1}_{i,\max}},
\end{equation}
and its optimal value is given by $H_{k,\max}\eta_k = 1/{\sum\nolimits_{i \in \cal K}H^{-1}_{i,\max}}, k \in {\cal K}$. It follows from (\ref{opt_eta}) that the optimal dynamic beam routing scheme can simply perform time sharing for all $K$ EUs, by assigning each of them a time slot for exclusive WPT.

Thus, the key to solving (P2.2) lies in how to obtain $H_{k,\max}, k \in \cal K$, which is the maximum channel power gain achievable by EU $k$. It can be obtained by solving the following single-user multi-path beam routing problem for EU $k$, i.e.,
\begin{align}\label{SUBR}
\mathop {\max}\limits_{\Gamma_k}\;&\sum\limits_{\Omega \in \Gamma_k}H_{0,J+k}(\Omega) \nonumber\\
\text{s.t.}\;\;&e(a^{(q)}_{k,l},a^{(q)}_{k,l+1}) \in E_L, \forall 0 \le l \le L_k^{(q)}, q \in {\cal Q}_k, \nonumber\\
&\Omega_k^{(q)} \cap \Omega_k^{(q')} =\emptyset, \forall q,q' \in {\cal Q}_k, q \ne q',
\end{align}
which is equivalent to solving (P1) with $K=1$ and EU $k$ assigned for WPT only (thus $\rho_k=1$).

A similar problem to (\ref{SUBR}) has been studied in our prior work\cite{mei2022split}, and we only outline the main steps of solving (\ref{SUBR}) next. Note that problem (\ref{SUBR}) can be optimally solved by enumerating all possible reflection path sets for EU $k$, i.e., $\Gamma_k$, and comparing their achieved channel power gains, which, however, may incur excessively high complexity. To reduce the enumeration complexity, we consider a clique-based partial enumeration approach. Specifically, we first select $U\;(U \ge 1)$ candidate paths which achieve the $U$ largest channel power gains from the BS to EU $k$ (i.e., $H_{0,J+k}(\Omega)$). This can be achieved by creating a line graph based on the LoS graph $G_L$ and applying the Yen's algorithm\cite{west1996introduction} to it. The details on how to create the line graph can be found in \cite{mei2022mbmh} and thus omitted here for brevity. Let ${\cal U}_k$ be the set of candidate reflection paths for EU $k, k \in \cal K$. For convenience, we denote by $P_k^{(u)}$ the $u$-th candidate path in ${\cal U}_k$, with $H_{0,J+k}(P_k^{(1)}) \ge H_{0,J+k}(P_k^{(2)}) \ge \cdots \ge H_{0,J+k}(P_k^{(U)})$. 

Next, among the $U$ paths in ${\cal U}_k$, we construct all possible reflection path sets for EU $k$ from them. To this end, we can construct a path graph $G_{p,k} = (V_{p,k},E_{p,k})$, where each vertex in $V_{p,k}$ corresponds to one candidate path, i.e., $V_{p,k} = \{v(P_k^{(u)})|u \in \{1,2,\cdots,U\}\}$, with $v(P_k^{(u)})$ denoting the vertex corresponding to the path $P_k^{(u)}$. To ensure that any two selected reflection paths are node-disjoint or satisfy the constraint in (\ref{SUBR}), we add an edge between any two vertices in $V_{p,k}$ (e.g., $v(P_k^{(u)})$ and $v(P_k^{(u')})$) if and only if their corresponding paths in $G_L$ are node-disjoint (i.e., $P_k^{(u)} \cap P_k^{(u')} = \emptyset$). Next, we present the definition of clique in graph theory.
\begin{definition}\label{clique}
A clique refers to a subset of vertices of an undirected graph such that every two distinct vertices in the clique are adjacent.	
\end{definition}

Based on Definition \ref{clique}, a clique with $S$ vertices in $G_{p,k}$ correspond to $S$ node-disjoint paths in ${\cal U}_k$. Hence, we can enumerate all cliques in $G_{p,k}$ to obtain all reflection path sets for EU $k$ from ${\cal U}_k$. However, there is in fact no need to enumerate all possible cliques in $G_{p,k}$. To validate this fact, we present the following definition.

\begin{definition}\label{Maxclique}
A maximal clique refers to a clique that cannot be extended by including one more adjacent vertex, that is, a clique which does not exist exclusively within the vertex set of a larger clique.
\end{definition}

It follows from Definition \ref{Maxclique} that we only need to enumerate the maximal cliques in $G_{p,k}$. This is because for any clique in $G_{p,k}$ (e.g., $C$), if it is not a maximal clique, there must exist another clique $C_0$ satisfying $C \subset C_0$. Then, the corresponding reflection paths of $C$ should also be a subset of those of $C_0$. Hence, the latter must yield a better performance than the former, as EU $k$ can be served by the BS over more paths. To enumerate all maximal cliques in a graph, we can invoke the Bron-Kerbosch algorithm, and the required worst-case computational complexity is in the order of $3^{U/3}$\cite{west1996introduction}. Let ${\cal C}_{k,\max}$ denote the set of all maximal cliques in $G_{p,k}$. For any maximal clique $C$ in ${\cal C}_{k,\max}$, let $C^{(k)}$ denote its corresponding reflection path set for EU $k$. Finally, denote by $\tilde H_{k,\max}$ the maximum channel power gains among all maximal cliques in ${\cal C}_{k,\max}$, i.e., \[\tilde H_{k,\max}=\mathop {\max}\limits_{C \in {\cal C}_{k,\max}}\;\sum\limits_{\Omega \in C^{(k)}}H_{0,J+k}(\Omega).\] After completing the above procedures for all $K$ EUs, the performance of the proposed partial enumeration approach can be obtained by replacing $H_{k,\max}$ in (\ref{opt_eta}) with $\tilde H_{k,\max}, k \in \cal K$. Evidently, by increasing $U$, more reflection path sets for each EU may be constructed, which helps improve $\tilde H_{k,\max}$. In particular, if $U$ is set to be sufficiently large, such that all possible reflection path sets for each EU can be constructed, then we can achieve $\tilde H_{k,\max}=H_{k,\max}, k \in \cal K$ and optimally solve (P1), but at the cost of more computational complexity.\vspace{-9pt}

\subsection{Proposed Solution to (P1)}
Although the optimal solution to (P1) can be derived by setting $T=1$ in (P2.1), the reflection paths for different EUs cannot be decoupled in a time-sharing manner as in (P2.2). To tackle this difficulty, we still first select a candidate path set ${\cal U}_k$ for each EU $k$. Let ${\cal U}=\bigcup\nolimits_{k=1}^{K} {\cal U}_k$ denote the union of the candidate path sets for all EUs. Then, we construct all possible path combinations for them or feasible solutions to (P1) from ${\cal U}$ by utilizing the clique-based approach. In particular, we create a path graph $G_p = (V_p,E_p)$ including all paths in ${\cal U}$ (instead of ${\cal U}_k$ as in Section \ref{solP2}), i.e., $V_p = \{v(P_k^{(u)})|k \in {\cal K}, u \in \{1,2,\cdots,U\}\}$. Next, we add an edge between any two vertices in $V_p$ (e.g., $v(P_k^{(u)})$ and $v(p_{k'}^{(u')})$) if and only if their corresponding paths in $G_L$ are node-disjoint (i.e., $P_k^{(u)} \cap P_{k'}^{(u')} = \emptyset$). It can be similarly shown that a clique in $G_p$ corresponds to one feasible solution to (P1), and it suffices to find the maximal cliques in $G_p$.

Let ${\cal C}_{\max}$ denote the set of all maximal cliques in $G_p$. Then, for any maximal clique $C$ in ${\cal C}_{\max}$, its corresponding channel power gain of EU $k$ is given by $E_k(C)=\sum\nolimits_{\Omega \in C^{(k)}}H_{0,J+k}(\Omega)$. In the case of $C^{(k)}=\emptyset$, we can set $E_k(C)=0$. Given $E_k(C), k \in \cal K$, we next optimize the power allocations for the EUs, i.e., $\rho_k, k \in \cal K$, to maximize $\mathop {\min}\nolimits_{k \in \cal K}\rho_kE_k(C)$ subject to $\sum\nolimits_{k \in \cal K}{\rho_k}=1$. Similar to (\ref{opt_eta}), the optimal value of this problem can be obtained as
\begin{equation}
	E(C) = \frac{1}{\sum\limits_{k \in \cal K}E^{-1}_k(C)},
\end{equation}
by setting $\rho_k=\frac{E^{-1}_k(C)}{\sum\nolimits_{i \in \cal K}E^{-1}_i(C)}$. Finally, we let $C^{\star}=\arg\mathop {\max}\nolimits_{C \in {\cal C}_{\max}}E(C)$ and the reflection path set for EU $k$ is optimized as $C^{\star(k)}$. Similar to Section \ref{solP2}, this clique-based approach may only find a suboptimal solution to (P1), which improves as $U$ becomes larger.

\section{Simulation Results}\label{sim}
\begin{figure}[!t]
\centering
\subfigure[3D plot.]{\includegraphics[width=0.37\textwidth]{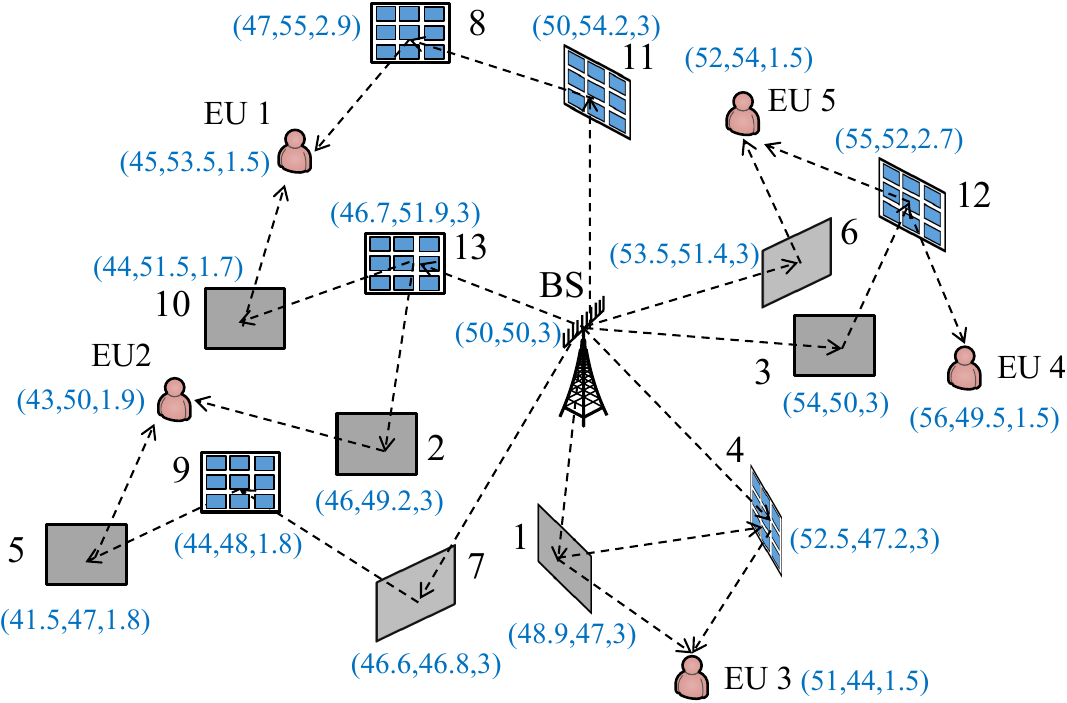}}
\subfigure[LoS graph.]{\includegraphics[width=0.37\textwidth]{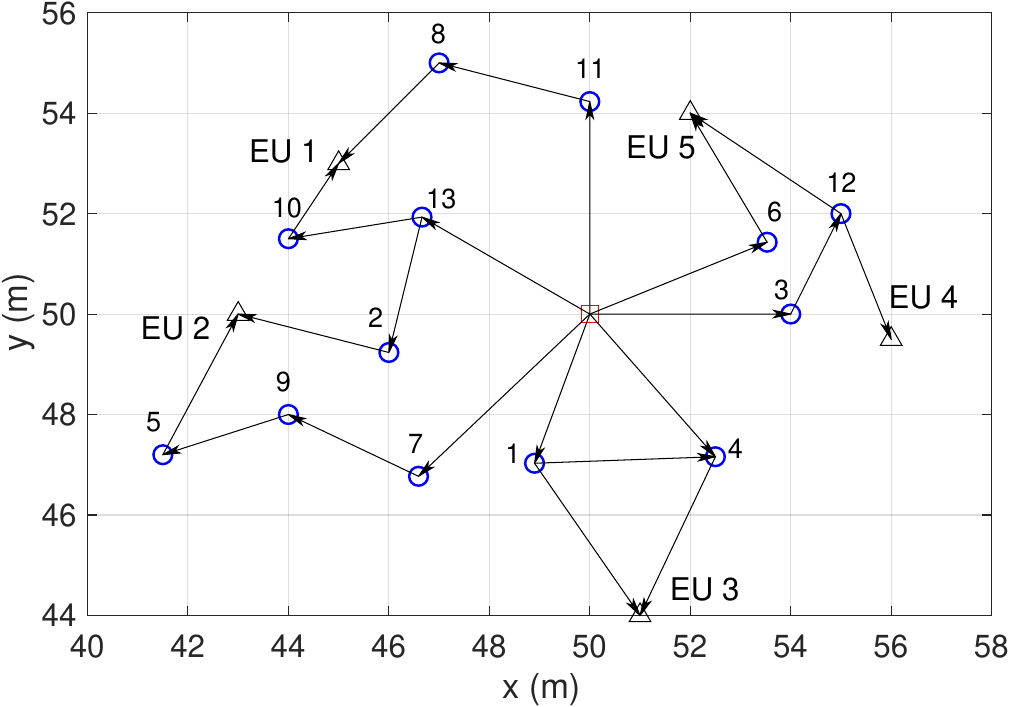}}
\vspace{-6pt}
\caption{Simulation setup of the multi-IRS-reflection WPT system.}\label{topology}
\vspace{-9pt}
\end{figure}
In this section, simulation results are provided to demonstrate the WPT performance of the proposed dynamic beam routing scheme. The three-dimensional (3D) coordinates of all nodes involved and the facing directions of all IRSs in this system are shown in Fig.\,\ref{topology}(a), while the corresponding LoS graph $G_L$ is shown in Fig.\,\ref{topology}(b). The numbers of IRSs and EUs are $J=13$ and $K=5$, respectively. The number of BS antennas is $N_B=32$ with $d_A=\lambda/2$. The carrier frequency is set to $f_c=5$ GHz with $\beta=-46$ dB. The numbers of reflecting elements in each IRS's horizontal and vertical dimensions are assumed to be identical as $M_0 \triangleq \sqrt{M}$, with $d_I=\lambda/4$. We consider that each IRS applies 3D passive beamforming and use 64-point discrete Fourier transform (DFT)-based codebooks for its horizontal and vertical passive beamforming, respectively; while the BS uses 32-point DFT-codebook for its active energy beamforming. The transmit power of the BS is set to be 30 dBm. We compare the WPT performance of the proposed dynamic beam routing with that of the baseline static beam routing. In addition, we also consider another benchmark scheme where the total transmission time is equally allocated for the $K$ EUs, i.e., setting $\eta_k=1/K, k \in \cal K$ in (P2.2).
\begin{figure*}[!t]
\centering
\subfigure[Static beam routing]{\includegraphics[width=0.42\textwidth]{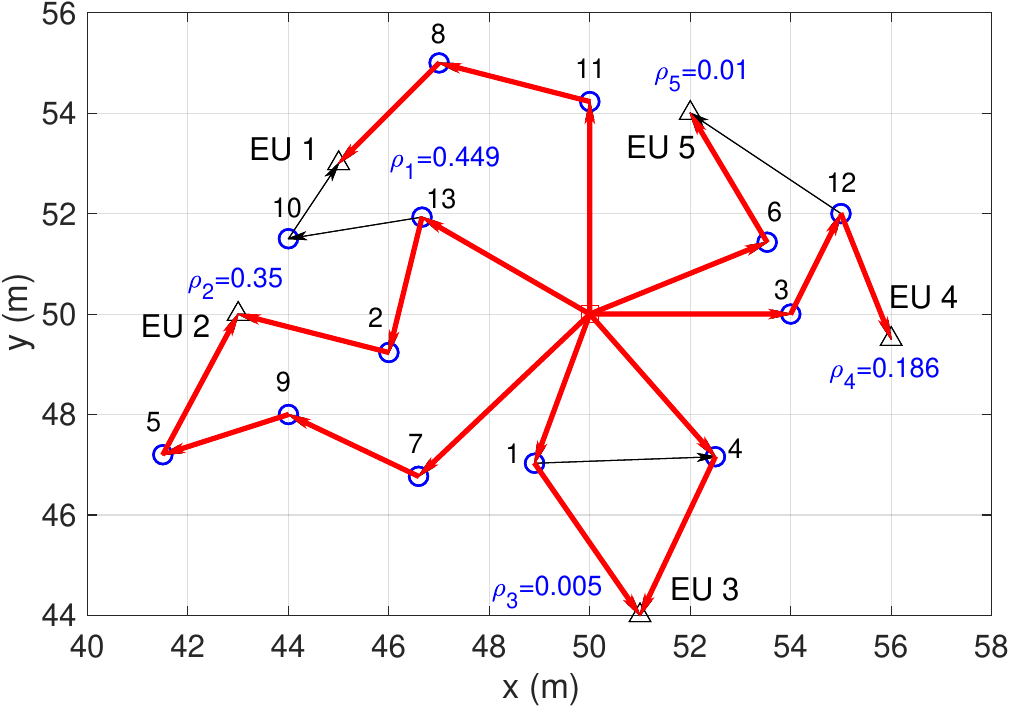}}
\subfigure[Dynamic beam routing, time slot 1]{\includegraphics[width=0.42\textwidth]{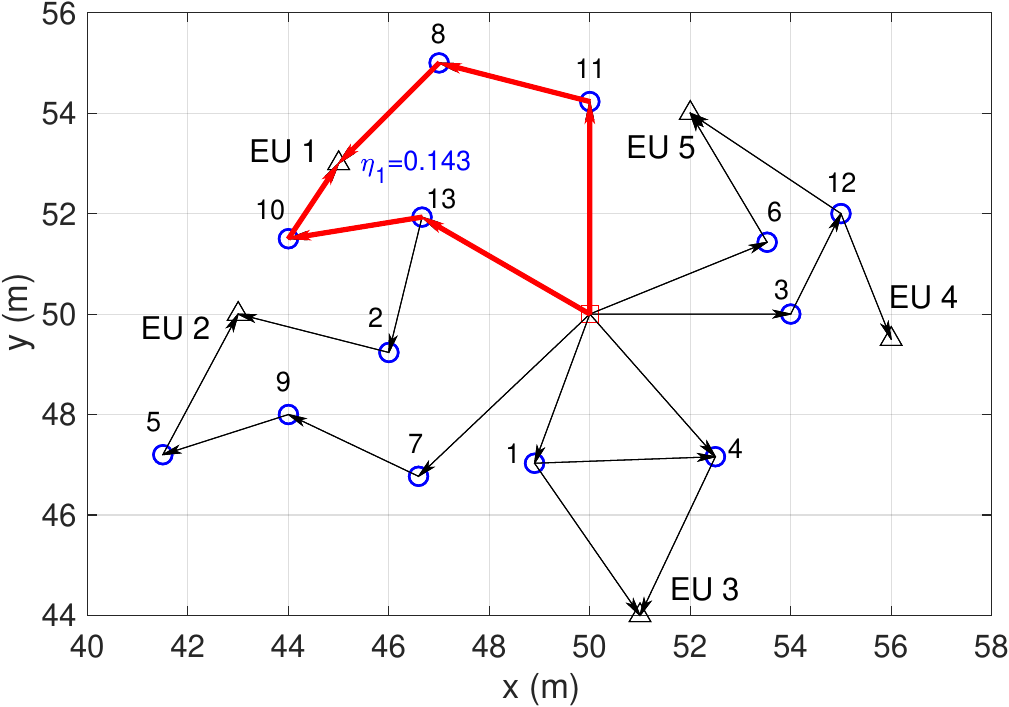}}
\subfigure[Dynamic beam routing, time slot 2]{\includegraphics[width=0.42\textwidth]{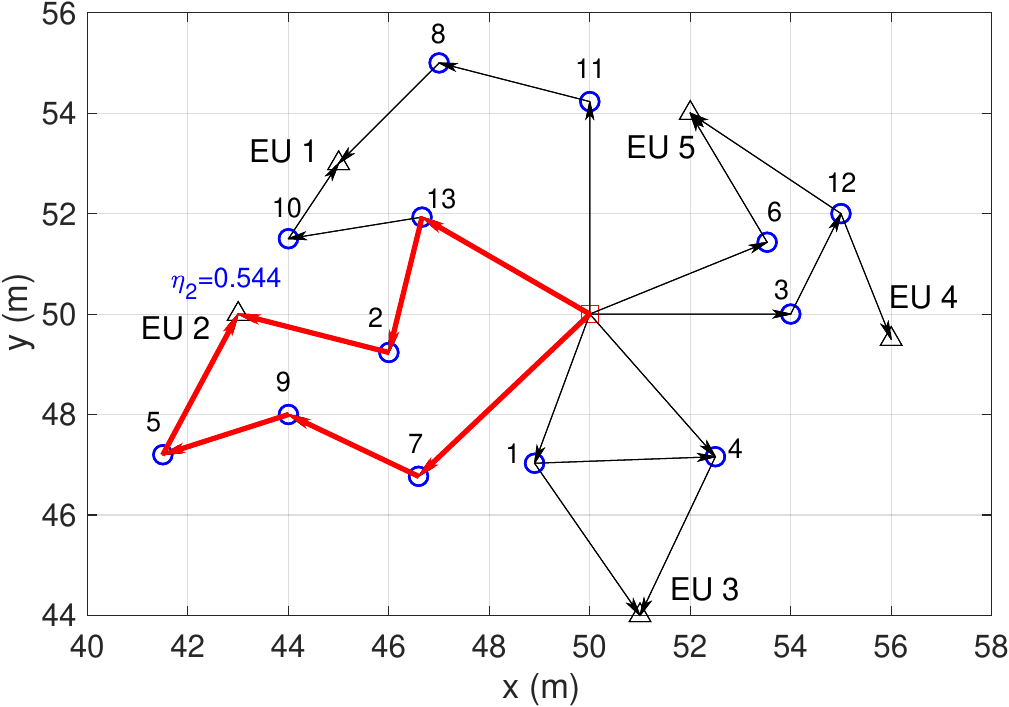}}
\subfigure[Dynamic beam routing, time slot 3]{\includegraphics[width=0.42\textwidth]{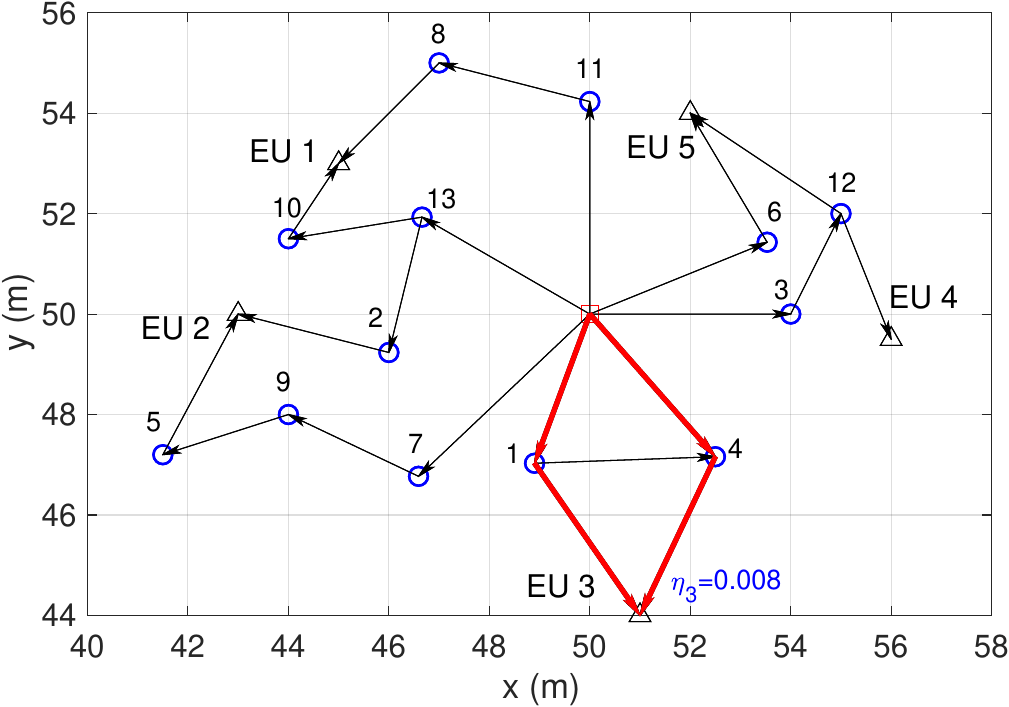}}
\subfigure[Dynamic beam routing, time slot 4]{\includegraphics[width=0.42\textwidth]{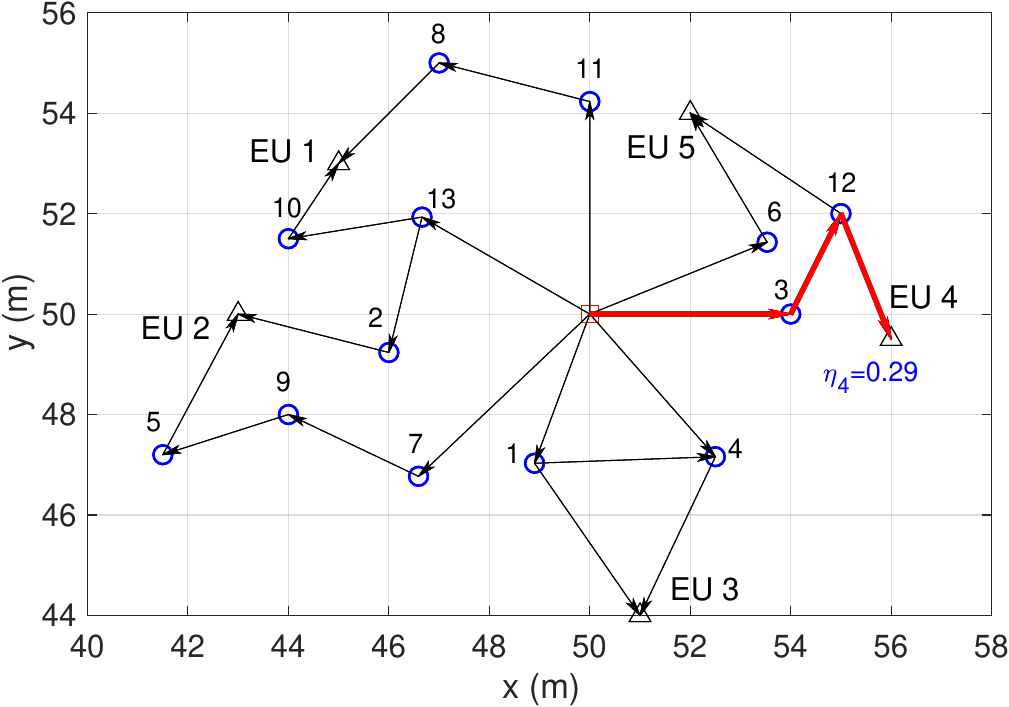}}
\subfigure[Dynamic beam routing, time slot 5]{\includegraphics[width=0.42\textwidth]{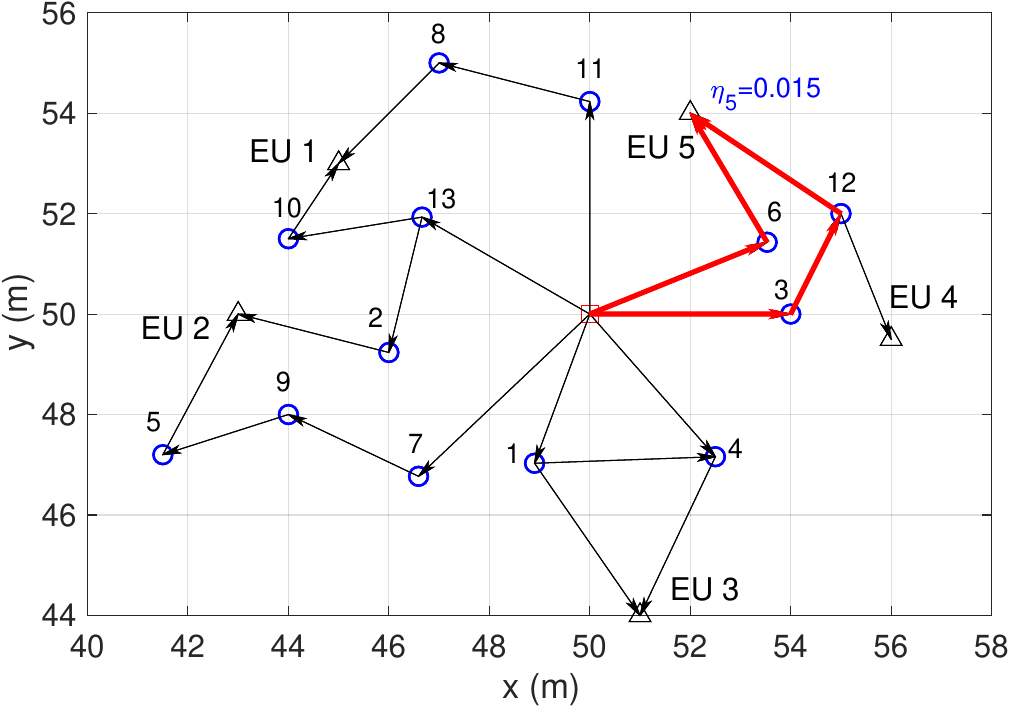}}
\caption{Optimized reflection paths and resource allocations by the static and dynamic beam routing schemes.}\label{RefPath}
\vspace{-6pt}
\end{figure*}

First, we plot the optimized reflection paths and resource allocations for all EUs by the static and dynamic beam routing schemes in Fig.\,\ref{RefPath}. It is observed from Fig.\,\ref{RefPath}(a) that all EUs can be assigned for WPT at the same time under the static beam routing scheme. In particular, less power is allocated to the EUs with higher end-to-end channel power gains with the BS, e.g., EU 3, which connects with the BS via two single-IRS-reflection links. However, due to the coupling of the reflection paths among different EUs, some EUs cannot achieve the maximum channel power gain with the BS. For example, both EUs 1 and 5 can be served by the BS over two IRS-reflection links; however, to assign EUs 2 and 4 for WPT, they only connect with the BS via a single IRS-reflection link. While in the proposed dynamic beam routing scheme, all EUs are assigned for WPT over different time slots in a time-sharing manner, such that they can achieve the maximum channel power gains with the BS in their respectively assigned time slots, as observed from Fig.\,\ref{RefPath}(b) to Fig.\,\ref{RefPath}(f). In particular, for EUs 2, 3 and 4, their optimized reflection paths are the same as those in Fig.\,\ref{RefPath}(a), but their allocated time durations (or powers) are higher than those in Fig.\,\ref{RefPath}(a), due to the improved channel power gains from the BS to EUs 1 and 5. It should also be mentioned that since the optimized reflection paths for EUs 2, 3, and 4 are node-disjoint in the proposed dynamic beam routing scheme, they can in fact be assigned into the same time slot (e.g., time slot 2) for WPT, which does not affect the received power but helps reduce the power transmission delay for EUs 3 and 4.

\begin{figure}[!t]
\centering
\includegraphics[width=3in]{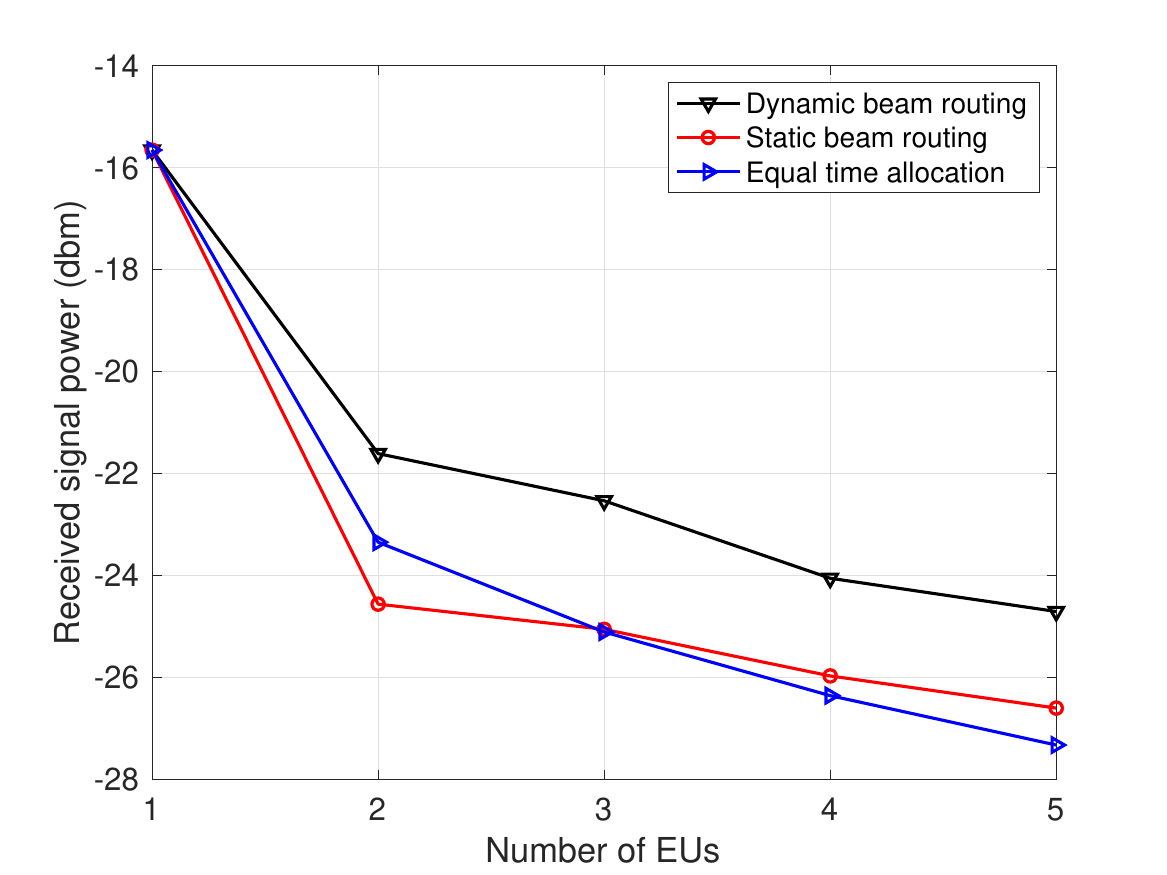}
\DeclareGraphicsExtensions.
\caption{Minimum received signal power versus the number of EUs.}\label{RecPwVsEUNum}
\vspace{-15pt}
\end{figure}
Next, we plot in Fig.\,\ref{RecPwVsEUNum} the minimum received power among all EUs by different schemes versus the number of EUs with $M_0=20$, where we assume that EUs $1, 2, \cdots, i$ are assigned for WPT when $K=i$. It is observed that with increasingly more EUs assigned for WPT, the overall performance degrades since the objective function of (P2.2) is non-increasing with $K$. In addition, when $K=1$ or only EU 1 is assigned for WPT, all considered schemes are observed to achieve the same performance, since EU 1 can achieve the maximum end-to-end channel power gain with the BS in any scheme in this case. However, when $K>1$, the proposed dynamic beam routing scheme can achieve much better performance than the other two benchmark schemes. The reason is that for the benchmark scheme with equal time allocations, it does not leverage the degree of freedom for optimizing the time allocations among all EUs. While for the static beam routing, due to the node-disjoint constraints on the reflection paths, the WPT performance of some EUs (e.g., EUs 1 and 5) has to be sacrificed to allow all EUs to perform WPT at the same time. This observation manifests that the dynamic beam routing scheme can achieve more significant performance gain over its static counterpart when the optimal reflection paths for different EUs have more shared IRSs (e.g., EU 1 and EU 2, as well as EU 4 and EU 5). This may happen more frequently when the number of EUs further increases, such that there is no feasible solution to (P1) and the performance of the static beam routing is virtually zero power in this case.

Finally, Fig.\,\ref{RecPwVsM} shows the minimum received power among all EUs by different schemes versus the number of reflecting elements per dimension, $M_0$, with $K=5$. It is observed that the WPT performance by all schemes monotonically increases with $M_0$, thanks to the more significant CPB gain achievable by each reflection path. In addition, the proposed dynamic beam routing outperforms the other two benchmark schemes over the whole range of $M_0$ considered. In particular, the performance gap between the proposed dynamic beam routing and its static counterpart is observed to become larger as $M_0$ increases. Particularly, when $M_0$ is sufficiently large (i.e., $M_0 \ge 22$), the benchmark scheme with equal time allocations also yields a better performance than the static beam routing. This is because the number of reflecting IRSs over the optimized reflection paths may become larger with increasing $M_0$, so as to achieve higher CPB gain, which in turn increases the number of shared IRS nodes among the reflection paths for different EUs, thus resulting in lower performance improvement of the static beam routing with increasing $M_0$ as compared to the other two schemes.
\begin{figure}[!t]
\centering
\includegraphics[width=3in]{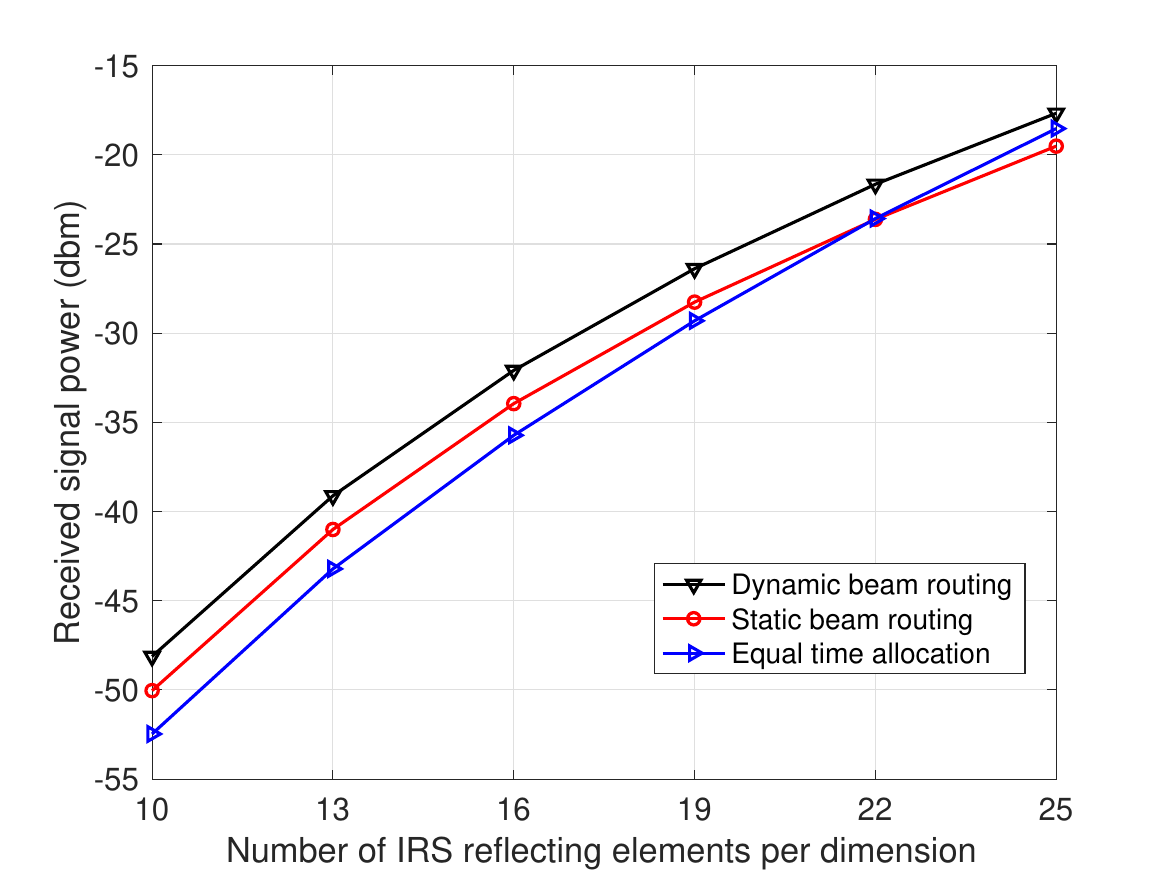}
\DeclareGraphicsExtensions.
\caption{Minimum received signal power versus the number of reflecting elements per dimension.}\label{RecPwVsM}
\vspace{-12pt}
\end{figure}

\section{Conclusions}
In this paper, we proposed two new beam routing schemes for multi-IRS-reflection aided WPT, namely, dynamic beam routing and subsurface-based beam routing, which enable each IRS to serve multiple EUs over different time slots and via different subsurfaces, respectively. In addition, we presented a baseline beam routing scheme with each IRS serving at most one EU, and proposed a new technique via adding random common IRS phase shift to tackle the undesired inter-path interference issue. Furthermore, it was shown that the dynamic beam routing outperforms the subsurface-based beam routing in terms of minimum harvested power among all EUs due to the more dominant effect of CPB gain over WPT duration. A clique-based optimization approach was also proposed to jointly optimize the beam routing and resource allocation in the proposed dynamic beam routing and baseline beam routing schemes. Finally, our numerical results verify the superior performance of the dynamic beam routing over the static beam routing, especially when the number of shared IRSs in the reflection paths for different EUs is large, even if they are not closely located. This paper can be extended along several directions in future work. For example, it is interesting to compare the performance of the proposed beam routing schemes in other scenarios, e.g., multi-carrier-based WPT and/or simultaneous wireless information and power transfer (SWIPT), for which the design needs to take into account the communication user performance\cite{wu2022intelligent}.\vspace{-9pt}

\bibliography{IRS_WPT.bib}
\bibliographystyle{IEEEtran}
\end{document}